\documentclass[11pt,x11names,a4paper]{article}

\usepackage{multirow}
\usepackage[utf8]{inputenc}
\usepackage{lmodern}
\usepackage[T1]{fontenc} 
\usepackage{microtype} 

\usepackage[a4paper, left=25mm, right=25mm, top=30mm, bottom=25mm]{geometry} 

\usepackage{abstract}

\usepackage{xcolor}

\usepackage{cite}
\usepackage{hyperref}
\hypersetup{
	colorlinks=true,
	linkcolor=Blue4,
	citecolor=Red4,
	urlcolor=Green4,
	linktoc=page
}

\usepackage{amsmath,amssymb,slashed,mathbbol,mathtools}
\numberwithin{equation}{section}

\usepackage{booktabs,overpic}

\newcommand{\dd}{\mathrm{d}}

\newcommand{\e}{\mathrm{e}}
\newcommand{\w}{\wedge}

\newcommand{\be}{\begin{equation}}
\newcommand{\ee}{\end{equation}}
\newcommand{\bea}{\begin{eqnarray}}
\newcommand{\eea}{\end{eqnarray}}

\newcommand{\f}[2]{\frac{#1}{#2}}
\newcommand{\R}{\mathbf{R}}

\renewcommand{\Re}{\text{Re}}
\renewcommand{\Im}{\text{Im}}

\newcommand{\Tr}{\text{Tr}~}

\newcommand{\U}{\text{U}}
\newcommand{\OO}{\text{O}}
\newcommand{\SU}{\text{SU}}
\newcommand{\SO}{\text{SO}}

\newcommand{\USp}{\text{USp}}
\newcommand{\SL}{\text{SL}}

\newcommand{\Ess}{\text{E}_{6(6)}}
\newcommand{\ess}{\mathfrak{e}_{6(6)}}

\newcommand{\Z}{\mathbf{Z}}

\title{\fontsize{20pt}{24pt}\selectfont\textbf{New analytic Janus and $S^4$ solutions for ${\cal N}=2^*$}\vspace{3mm}}

\author{
\large{\href{mailto:ffg@soton.ac.uk}{Fri{\dh}rik Freyr Gautason}}\\[5mm]
{STAG Research Centre $\&$ Mathematical Sciences, University of Southampton,}\\ 
{\normalsize University Road, Southampton SO17 1BJ, UK}\\[5mm]
}

\begin{document}
{\hypersetup{urlcolor=black}\maketitle}
\thispagestyle{empty}

\begin{abstract}
\noindent We construct a three-parameter family of regular analytic supergravity solutions which are dual to (a) the $\mathcal{N}=2^*$ gauge theory on the four-sphere, and (b) $\U(1)^2$-invariant Janus interfaces in $\mathcal{N}=4$ SYM with or without the $\mathcal{N}=2^*$ mass turned on. To this end we employ a five-dimensional supergravity model coupled to  six scalars which arises as a certain consistent truncation of type IIB supergravity on $S^5$. The curved domain wall solutions feature a hidden integral of motion which allows for their analytic treatment. We relate the three parameters of the solution to field theory sources and vevs and show how regularity of the solutions restricts their range. We briefly discuss the holographic evaluation of the one-point function of the field theory Lagrangian for our family of Janus solutions.

\end{abstract}

\newpage
\tableofcontents

\section{Introduction}

From the early days of holography, gauged supergravity has played a pivotal role in providing examples of dual pairs between quantum field theories and string backgrounds. Five-dimensional gauged supergravity \cite{Gunaydin:1985cu,Pernici:1985ju} arises as a consistent truncation of type IIB supergravity on $S^5$ \cite{Lee:2014mla,Baguet:2015sma} and can be used as a simplifying tool to find solutions of type IIB supergravity without having to solve Einstein's equations in ten dimensions. It thus serves as an ideal playground to explore the vacuum structure of the dual field theory \cite{Freedman:1999gk,Cvetic:1999xx} and its deformations whether they have strongly coupled IR fixed points \cite{Khavaev:1998fb,Girardello:1998pd,Freedman:1999gp,Krishnan:2020sfg,Bobev:2020ttg} or not \cite{Girardello:1999bd,Pilch:2000ue}. Gauged supergravity can be further used to make holographic predictions which can be tested against other, more direct field theory computations. This story has been revitalized with the advent of supersymmetric localization \cite{Pestun:2007rz} which gives direct access to a subset of observables in strongly coupled supersymmetric QFTs and thus enables the verification of supergravity results. 

A direct comparison with results from supersymmetric localization often requires the study of domain wall solutions of supergravity where the wall is curved. A prime example is the mass-deformed $\mathcal{N}=4$ Supersymmetric Yang--Mills (SYM) on $S^4$. In $\mathcal{N}=1$ language the $\mathcal{N}=4$ theory consists of a vector multiplet together with three chiral multiplets and a cubic superpotential. All fields transform in the adjoint of the $\SU(N)$ gauge group. Introducing independent mass terms for the chiral multiplets results in a genuine $\mathcal{N}=1$ gauge theory but supersymmetry may be enhanced by equating two out of three masses and setting the third one to zero. This results in the so-called $\mathcal{N}=2^*$ theory which can be localized on the four-sphere. The $\mathcal{N}=2^*$ theory is not conformal and in order to preserve supersymmetry on the sphere, additional couplings must be switched on. On the gravity side, this also means that the holographic dual is only asymptotically locally AdS$_5$ but is otherwise deformed.

Even though the type IIB theory can be consistently truncated to five dimensions, it does not automatically follow that the holographic dual we are after is a solution of the truncated theory. A zeroth-order requirement is that the deformation only turns on operators in the stress-tensor multiplet whose dual is the gravity multiplet of the five-dimensional theory. If other multiplets are turned on in the QFT, then we must construct the dual directly in ten dimensions or in some larger theory which has access to these operators (although such a theory will most likely not be a consistent truncation). This is further complicated by the fact that along the RG flow heavier operators may be turned on if symmetries allow for it and hence even if perturbative boundary analysis indicates that a particular construction should find a dual in five dimensions, it is by no means guaranteed to be regular. If we assume that the holographic dual is captured by the five-dimensional consistent truncation and that the entire flow does preserve the same symmetry as the boundary analysis implies (and it is not spontaneously broken), we may use that same symmetry to truncate the gauged supergravity theory to a smaller subsector thus simplifying the task of finding the dual gravitational solution.

In this paper we explore deformations of $\mathcal{N}=4$ SYM which can be described in holography as regular gravitational solutions of five-dimensional gauged supergravity. Our interest will be on two setups which a priori seem unrelated but are connected by the symmetry they preserve, and as a result their holographic description. The first, which we have already mentioned, is the $\mathcal{N}=2^*$ theory on $S^4$. Its holographic description was originally considered in \cite{Bobev:2013cja} where the gravitational solution was constructed numerically. Previously, the holographic dual to $\mathcal{N}=2^*$ on flat space had been constructed \cite{Pilch:2000ue} but that solution is singular, terminating in the so-called Enhan\c{c}on \cite{Johnson:1999qt,Buchel:2000cn,Evans:2000ct}. Spherical domain walls have the benefit that the singularity can be excised and holographic interpretation of the solution is straightforward. Using careful numerical analysis of the BPS equations, the authors of \cite{Bobev:2013cja} were able to find a match with supersymmetric localization for the (scheme-independent part of the) sphere free energy \cite{Buchel:2013id} by computing the renormalized five-dimensional on-shell action. In this work we construct the \emph{analytic} domain wall solution, recovering the relation between the mass and the vev of a dimension-two operator found numerically in \cite{Bobev:2013cja}. The match with localization relied on this relation which therefore carries over.

Another class of setups we consider are conformal interfaces in $\mathcal{N}=4$ SYM with and without the $\mathcal{N}=2^*$ mass turned on. These interfaces are characterized by the jump of the complexified coupling constant across the interface \cite{DHoker:2006qeo,Gaiotto:2008sd} and by a mass which varies as a function of the distance from it \cite{Arav:2020obl}. The holographic dual description is in terms of AdS$_4$ sliced domain walls preserving 3D $\mathcal{N}=2$ supersymmetry with two asymptotic AdS$_5$ regions and a curved geometry that connects them. The AdS$_4$ slices are motivated by the fact that conformal interfaces preserve the 3D conformal group which is represented in gravity as the isometry group of AdS$_4$. Such backgrounds are often denoted supersymmetric Janus solutions due to their two-faced nature \cite{Bak:2003jk,Clark:2004sb}. Constructions of AdS$_5$ Janus solutions in gravity have been carried out both in five dimensions (and then often uplifted to ten dimensions) \cite{Clark:2005te,Suh:2011xc,Bobev:2019jbi,Bobev:2020fon,Arav:2020asu,Arav:2020obl} as well as directly in ten dimensions \cite{DHoker:2006vfr,DHoker:2007zhm,DHoker:2007hhe}. In this paper we construct a three-parameter family of analytic Janus solutions which are parametrized by the $\mathcal{N}=2^*$ mass $\mu$, and two other constants $\mathcal{J}$ and $\gamma$ which control the jump of the dilaton across the Janus, and a vev of a dimension-two operator respectively. Only two one-dimensional subfamilies were known in closed form before this work, both of which have enhanced symmetry and were constructed in \cite{Bobev:2020fon}. A two-parameter family of massive Janus solutions of the three-scalar model of \cite{Bobev:2013cja} was analyzed numerically in \cite{Arav:2020obl}. It is contained in our family which therefore provides these solutions in closed form.

The solutions we construct are controlled by a classical scattering of a particle off an inverted harmonic oscillator (see Figure~\ref{fig:potential} below). The particle position is directly related to a metric function that controls the radius of the AdS$_4$ slice. For the massless solutions $\mu=0$, when the energy of the particle is tuned to the maximum of the potential, a different class of solutions emerges where the particle stays put at the top of the hill. The metric and all scalar fields except one are constant. Only the dilaton is linear in the domain wall coordinate $r$ which ordinarily would spell doom for the solution as the string coupling runs to large values. However in \cite{Assel:2018vtq} it was understood how this can be remedied by compactifying the radial direction along which the dilaton runs at the same time as a duality wall for the $\SL(2,\Z)$ symmetry of type IIB string theory is introduced. These solutions are called J-folds and are holographic duals to 3D SCFTs which have a similar close relation to interfaces in the 4D parent theory. The family of five-dimensional J-folds which we recover was constructed in \cite{Bobev:2020fon,Arav:2021tpk,Arav:2021gra}.

As hinted above, both the $\mathcal{N}=2^*$ mass and the interfaces we consider preserve $\U(1)^2$ in addition to $\Z_2^2$ and imposing that symmetry on the maximal supergravity in five dimensions leads to a tractable six-scalar model which we focus on. We consider domain wall solutions where the five-dimensional metric takes the form
\be
\dd s_5^2 = \dd r^2 + \e^{2A}\dd s_{4}^2\,,
\ee
where $\dd s_{4}^2$ is a metric on a maximally symmetric four-dimensional slice with curvature set by $-\kappa^2$. In this work we focus mainly on Lorentzian AdS$_4$ with $\kappa^2=+1$ and Euclidean $S^4$ with $\kappa^2=-1$. Flat slices with $\kappa^2=0$ behave differently and will be discussed elsewhere \cite{futurework}. The BPS equations reduce the equations of motion for $A$ and the scalar fields to first order differential equations for five functions. The six scalars include the five-dimensional dilaton which does not appear in the potential. This has two consequences: first, as the shift of the dilaton is a symmetry of the model, we can identify a Noether charge which implies an integral of motion. Second, the BPS equation for the dilaton itself decouples from the system and can be solved by quadrature. The key step to solving the BPS equations comes from identifying a \emph{hidden} integral of motion which can be understood as the $\mathcal{N}=2^*$ mass built out of the supergravity fields and extended into the bulk. This integral of motion is not a consequence of the symmetry of the theory, but rather the fact that two different combinations of scalar fields satisfy the same linear second-order equation: $\partial_s^2\ell_\pm = \ell_\pm$. From this we build the aforementioned integral of motion and derive the scattering problem that controls the solution.

The rest of the paper is organized as follows: in Sections~\ref{6scalarmodel} and \ref{sec:solution} we review the six-scalar model and find a general solution to the curved space domain wall BPS equations. In Section~\ref{sec:janus} we specify to AdS$_4$ slices and discuss Janus solutions. In particular we find a three-parameter family of solutions with or without the $\mathcal{N}=2^*$ mass turned on. We also briefly discuss the J-fold solutions in this model and extract the one-point function of the field theory Lagrangian and compare with recent work of \cite{Karch:2026ymg,He:2026qnv}. In Section~\ref{sec:sphere} we focus on spherical domain walls, and show how the general solution reduces by regularity to a one-parameter family of solutions controlled only by the $\mathcal{N}=2^*$ mass. We also briefly review the uplift of the sphere solution to ten dimensions carried out originally in \cite{Bobev:2018hbq}. We conclude with a discussion in Section~\ref{sec:discussion}. We include one appendix which relates the six-scalar model to the ten-scalar model of \cite{Bobev:2016nua}.

\section{The six-scalar model}\label{6scalarmodel}
In this paper we consider a six-scalar truncation of the maximal 5D supergravity with $\SO(6)$ gauge group \cite{Gunaydin:1985cu,Pernici:1985ju}. The maximal supergravity theory arises as a consistent truncation of type IIB supergravity on $S^5$ and as such any solution of the five-dimensional theory uplifts to a solution of type IIB supergravity \cite{Lee:2014mla,Baguet:2015sma}. As an example, the maximally supersymmetric AdS$_5$ vacuum solution of the five-dimensional theory uplifts to the well-known AdS$_5\times S^5$ near-horizon geometry of a stack of many D3-branes in ten-dimensional flat space. This five-dimensional perspective is therefore ideal to study holographic backgrounds which arise from deforming the dual ${\cal N}=4$ SYM theory by relevant operators and/or placing it on curved space.

The entire theory is however too large to meaningfully identify or even classify solutions. In addition to the metric we have 42 scalar fields, 15 vectors, 12 two-forms, and a wealth of fermions. The standard approach is to identify a suitable subset of fields which can be retained while all others are set to zero. This approach was taken in \cite{Bobev:2016nua} where the holographic dual to ${\cal N}=1^*$ on $S^4$ was considered. Here ${\cal N}=1^*$ is a commonly given name to the QFT that arises when the ${\cal N}=4$ SYM is deformed by supersymmetric mass parameters. There are three such mass terms, but when placing the theory on $S^4$ more operators must be turned on to preserve supersymmetry. The careful study of these operators led the authors of \cite{Bobev:2016nua} to consider a ten-scalar truncation of the maximal 5D supergravity. 

Here we focus on a further truncation of the ten-scalar model also discussed in \cite{Bobev:2016nua} that preserves $\U(1)^2$ continuous symmetry. This $\U(1)^2$ is embedded in $\SO(6)$ as the rotation of two 2-planes in $\R^6$ that $\SO(6)$ acts on. Demanding invariance with respect to $\U(1)^2\times \Z_2^2$ leaves six scalar fields which parametrize the manifold
\be\label{6scalarcoset}
\OO(1,1)^2\times\bigg[ \frac{\SL(2,\R)}{\U(1)} \bigg]^2\,.
\ee 
We refer to Appendix \ref{App:10and6scalars} for the details of how the six-scalar model embeds in the ten-scalar model of \cite{Bobev:2016nua}. Imposing the same $\U(1)^2$ symmetry on the boundary theory forces two of the three mass parameters in ${\cal N}=1^*$ to be identified and the third one to be set to zero. This is precisely how one obtains the ${\cal N}=2^*$ theory and hence the six-scalar model is an extension of the supergravity models in \cite{Pilch:2000ue,Bobev:2013cja} which were used to study ${\cal N}=2^*$ on $\R^4$ and $S^4$. The benefit of the six-scalar model is that it allows us to also consider a large class of interfaces dual to supersymmetric Janus solutions in ${\cal N}=4$ with or without a mass term turned on. The main reason for our study of this model however is that it turns out to be integrable and we can completely classify its domain wall solutions with AdS$_4$ or $S^4$ slices.

We emphasize that the six-scalar model is not a supergravity theory. The symmetries we have imposed eliminate all gravitini from the spectrum. This is not a problem, however, as we can still use supercharges of the maximal theory for the usual task of deriving BPS equations which reduce the problem of solving second-order equations of motion to first order.

The action for the six-scalar model is
\be
S = \f{1}{16\pi G_N} \int \star\Big(R-12 (\dd\alpha)^2-(\dd t)^2-2 K_{i\bar\jmath}\dd z^i\cdot\dd {\bar z}^{\bar\jmath}-{\cal P}\Big)\,,
\ee
where $t$ and $\alpha$ are real scalar fields whereas $z^i$ with $i=1,3$ denote two complex scalar fields. The esoteric choice of names for the two complex scalar fields is inherited from the ten-scalar model reviewed in Appendix \ref{App:10and6scalars}. Even though this theory is not a supergravity theory, we can still write the kinetic terms and the scalar potential in terms of a K{\"a}hler potential $K$ and a superpotential $\mathcal{W}$ as follows:
\be\label{Kaehlerandsuperpot}
\begin{split}
K &= - \log(2\Im z_1)- \log(2\Im z_3)\,, \qquad K_{i\bar\jmath} = \partial_i\partial_{\bar\jmath}K\,,\\
\mathcal{W} &=  \f{g}{2}\Big(\e^{-4\alpha}(\e^{-t}z_1z_3-\e^{t})+2i\e^{2\alpha}(z_1+z_3)\Big)\,,\\
{\cal P} &= \f12 \e^{K}\Big( \f16|\partial_\alpha \mathcal{W}|^2+2 |\partial_t \mathcal{W}|^2 + K^{i\bar\jmath}D_i \mathcal{W} D_{\bar\jmath}\overline{\mathcal{W}} - \f83 |\mathcal{W}|^2\Big)\,.
\end{split}
\ee
Here we have also utilized the K{\"a}hler covariant derivative $D_i\mathcal{W} = \partial_i\mathcal{W} + (\partial_iK)\mathcal{W}$ and $K^{i\bar\jmath}$ is the matrix inverse of $K_{i\bar\jmath}$. The supergravity gauge coupling constant $g$ is related to the length scale of the maximally supersymmetric vacuum AdS$_5$ via $L=2/g$. In holography, this is further related to the 't~Hooft coupling constant $\lambda$ of the dual gauge theory via $L^4/\ell_s^4 = \lambda$ where $\ell_s$ is the string length.

The superpotential (and its complex conjugate) is used to write supersymmetry variations for a pair of Killing spinors which we will use below to derive first order BPS equations. These Killing spinor variations can be reduced from their ten-scalar avatar presented in \cite{Bobev:2016nua} (see also Appendix \ref{App:10and6scalars}) and are
\be\label{spinorvar}
\begin{split}
\nabla_\mu \epsilon_1 - \f{i}4 Q_\mu \epsilon_1 - \f{i}6 \e^{K/2}{\overline{\mathcal{W}} }\gamma_\mu \epsilon_2 &= 0\,,\\
(\slashed \partial z^i) \epsilon_1 + \f{i}2 \e^{K/2}K^{i\bar\jmath}(D_{\bar\jmath} \overline{\mathcal{W}})\epsilon_2 &=0\,,\\
3(\slashed \partial \alpha) \epsilon_1 + \f{i}4 \e^{K/2}(\partial_\alpha \overline{\mathcal{W}})\epsilon_2 &=0\,,\\
(\slashed \partial t) \epsilon_1 + i \e^{K/2}(\partial_t \overline{\mathcal{W}})\epsilon_2 &=0\,,\\
\end{split}
\ee
where
\be
Q_\mu = -i(\partial_\mu z^i\partial_i-\partial_\mu{\bar z}^{\bar\imath}\partial_{\bar\imath})K\,.
\ee

\subsection{Scaling symmetry}
The six-scalar model exhibits a scaling symmetry
\be\label{scalingsymmetry}
z_i \mapsto \lambda z_i\,,\quad t\mapsto t+ \log(\lambda)\,,\quad K\mapsto K-2\log(\lambda)\,,\quad K_{i\bar\jmath}\mapsto \lambda^{-2}K_{i\bar\jmath}\,,\quad \mathcal{W}\mapsto \lambda \mathcal{W}\,,
\ee
which leaves the action invariant. This is nothing but the dilaton shift symmetry. Recall that two out of the 42 scalar fields in the maximal supergravity theory are a direct descendant of the axion-dilaton pair in type IIB supergravity and as such are subject to the same $\SL(2,\R)$ symmetry as their ten-dimensional avatar. The 5D dilaton is a part of the six-scalar model but not the axion and hence we only retain the non-compact part of $\SL(2,\R)$, namely the dilatation \eqref{scalingsymmetry}. This symmetry implies a conserved current
\be\label{Jformandconservation}
J= K_{i\bar\jmath} \dd (z^i{\bar z}^{\bar\jmath})+\dd t\,,\quad \dd\star J =\tfrac12\star\big(z^i\partial_i\mathcal P+\bar z^{\bar\jmath}\partial_{\bar\jmath}\mathcal P+\partial_t\mathcal P\big)=0\,.
\ee

\subsection{Relation to operators and UV expansion}
The scalar fields $z_i$ and $t$ are a mixture of the dilaton, ${\bf 20}'$ and ${\bf 10}\oplus\overline{\bf 10}$ scalars of the maximal theory. These in turn are dual to operators in the stress-tensor multiplet of the boundary gauge theory and it is therefore convenient to find a direct decomposition of the complex scalars to the so-called physical scalars which have a clear dual in the field theory. This was done for the ten-scalar model in \cite{Bobev:2016nua} and is repeated in Appendix \ref{App:10and6scalars} in our formulation. For convenience, it is useful to sketch the physical interpretation of the ten scalar fields of the parent model before specifying to the six-scalar model. In the ten-scalar model we have $\alpha$, $\beta$ and $\alpha_{1,2,3}$ which are dual to scalar bilinears and $\phi_{1,2,3}$ which are dual to fermion bilinear operators. In the ${\cal N}=1^*$ theory, the operators dual to $\phi_{1,2,3}$ are turned on which gives rise to masses for the three chiral multiplets. The supergravity scalar $\phi_4$ is dual to the gaugino bilinear and the 5D dilaton $\varphi$ is dual to the Yang--Mills Lagrangian itself. In the six-scalar model we have identified scalar fields and set some of them to zero, i.e. 
\be
\alpha_1 = \alpha_2\,,\qquad \phi_1 = \phi_2\,,\qquad \phi_3 = -\phi_4, \qquad \beta=0\,.
\ee
As we have already hinted, the interpretation is that supersymmetric solutions of the six-scalar model are dual to supersymmetric configurations of $\mathcal{N}=4$ SYM deformed by the ${\cal N}=2^*$ mass on curved space or in the presence of an interface. In ${\cal N}=2$ language, the ${\cal N}=4$ theory consists of a vector multiplet together with a hypermultiplet which for a general solution of our supergravity theory is given a mass $\mu$. Explicitly the physical scalars are related to the supergravity fields in the six-scalar model as follows
\be\label{physicalscalars}
\begin{split}
\log (-i z_1) &= 2\alpha_1+\alpha_3+\varphi - 2i\phi_1  - 2i\phi_3 \,,\\
\log (-i z_3) &= -2\alpha_1+\alpha_3+\varphi + 2i\phi_1  - 2i\phi_3 \,,\\
{t} &= -\alpha_3+\varphi  \,,\\
\end{split}
\ee
where we have used the same notation for physical scalars as in \cite{Bobev:2016nua}.

\section{Curved domain walls}\label{sec:solution}
We can derive the BPS equations for a curved domain wall with maximally symmetric slices compatible with $\SO(2,3)$ or $\SO(5)$ invariant metric depending on whether the slices are AdS$_4$ or $S^4$. Flat slices can be treated similarly but we will leave the discussion of flat space domain walls to future work \cite{futurework}. We write
\be
\dd s_5^2 = \dd r^2 + \e^{2A}\dd s_{4}^2\,,
\ee
where the metric function $A$ as well as all scalar fields are assumed to only depend on the variable $r$. To process the supersymmetry variations \eqref{spinorvar}, we assume the projector
\be
\gamma_{\hat r}\epsilon_2 = \e^{-i\xi} \epsilon_1\,,\qquad\gamma_{\hat r}\epsilon_1 = \e^{i\xi} \epsilon_2\,,
\ee
where hatted indices denote the flat (tangent space) indices and $\xi$ is some phase angle which depends on the radial coordinate $r$. Furthermore, we assume that the five-dimensional spinors are proportional to a (conformal) Killing spinor on the four-dimensional slice, meaning it satisfies
\be
\nabla_m \epsilon_1 = \f12 (\kappa i\e^{-A}-  A')\gamma_{\hat r}\gamma_m\epsilon_1\,,
\ee
and similar for $\epsilon_2$. Here $m$ is an index along the four-dimensional slices and $\kappa^2$ is $- 1$ for positively curved slice, $+1$ for negatively curved slices (and $0$ for flat slices). The resulting BPS equations are
\be\label{BPS6scalars}
\begin{split}
 A'-\kappa i\e^{-A} - \f{i}3 \e^{-i\xi}\e^{K/2}{\overline{\mathcal{W}} }  &= 0\,,\\
2(z^i)' + i\e^{-i\xi} \e^{K/2}K^{i\bar\jmath}(D_{\bar\jmath} \overline{\mathcal{W}}) &=0\,,\\
12\alpha'  + i\e^{-i\xi} \e^{K/2}(\partial_\alpha \overline{\mathcal{W}})&=0\,,\\
t' + i\e^{-i\xi} \e^{K/2}(\partial_t \overline{\mathcal{W}}) &=0\,.\\
\end{split}
\ee

Let us define new variables which will be useful in our analysis which follows. We let
\be\label{sigmadef}
\sigma_i = \e^{3A}\f{\Re z_i}{\Im z_i}\,,\quad \text{where}\quad i=1,3\,,
\ee
where the factor of $\e^{3A}$ is not chosen at random. It is directly related to the factor 1/3 in the $A$ BPS equation in \eqref{BPS6scalars}. These real functions satisfy the BPS equations
\be\label{sigmaBPS}
\sigma_i' = \e^{3A}\e^{K/2}\Re\Big[\e^{i\xi}(2 V_i-1)\mathcal{W}\Big]\,,
\ee
where $V_i = z^i\partial_i$ with no sum over $i$. We can now use this to simplify the conserved quantity discussed above. Let us first define
\be\label{conservedquantitydef}
{\cal J}\equiv\f{g^3}{12}\e^{4A}J_r\,,\qquad \partial_r {\cal J}=0\,.
\ee
The conservation follows from \eqref{Jformandconservation}, but we will now explicitly show this. First note that by the scaling symmetry, the superpotential exhibits a homogeneity property $(z^i\partial_i + \partial_t)\mathcal W = \mathcal W$. Using the vector fields $V_i$ we can also write this as $(V_1+V_3+\partial_t)\mathcal W=\mathcal W$. Furthermore, we have $2V_i K = -1+i\e^{-3A}\sigma_i$.
These together imply that
\be
(z^iD_i + \partial_t)\mathcal W = (1+V_1 K + V_3 K)\mathcal W = \f{i\e^{-3A}\sigma}{2}\mathcal W\,,
\ee
where $\sigma= \sigma_1+\sigma_3$. Now feed these relations into the definition of $J_r$ \eqref{Jformandconservation} using the BPS equations \eqref{BPS6scalars}
\be
J_r=K_{i\bar\jmath} \partial_r (z^i{\bar z}^{\bar\jmath})+\partial_r t
= -\f{\sigma}{2}\e^{-3A}\Re\Big[ \e^{-i\xi}\e^{K/2}{\overline{\mathcal{W}} } \Big]=\f{3\kappa\sigma}{2} \e^{-4A}\,,
\ee
where in the last step we used the imaginary part of the $A$ BPS equation in \eqref{BPS6scalars}. Combining with $\e^{4A}$ as in \eqref{conservedquantitydef}, we find that
\be\label{Jresult1}
{\cal J}=\f{\kappa g^3\sigma}{8} \,.
\ee
An alternative form can also be derived for the conserved quantity where it takes the following form
\be\label{Jresult2}
{\cal J}=\kappa^2 g \e^{2A+8\alpha}t'\,.
\ee

Let us now explicitly verify that \eqref{Jresult1} is indeed conserved. Using \eqref{sigmaBPS} we find that
\be
\partial_r{\cal J}=\f{\kappa g^3}{4}\e^{3A}\e^{K/2}\Re\Big[\e^{i\xi}(V_1+V_3-1)\mathcal{W}\Big] = -\f{\kappa g^3}{4}\e^{3A}\e^{K/2}\Re\Big[\e^{i\xi}(\partial_t\mathcal{W})\Big] =0\,,
\ee
where the last equality follows from the imaginary part of the $t$-BPS equation in \eqref{BPS6scalars}.

The BPS equations can be processed further without any loss of generality. In particular we can directly read off three algebraic equations
\be
-3\kappa \e^{-A}= \Re\Big[ \e^{i\xi}\e^{K/2}\mathcal{W} \Big]\,,\qquad
\Re\Big[\e^{i\xi} (\partial_\alpha \mathcal{W})\Big]=0=
\Re\Big[\e^{i\xi} (\partial_t \mathcal{W})\Big]\,.
\ee 
Using the explicit form of the superpotential \eqref{Kaehlerandsuperpot}, these can be combined and simplified to
\be\label{BPS6processed}
\begin{split}
\Re (\e^{i\xi}z_1z_3) + \e^{2t}\cos\xi&=0\,,\\
\Im \Big(\e^{i\xi}(z_1+z_3)\Big)-2\e^{t-6\alpha}\cos\xi&=0\,,\\
\kappa \e^{-A}- g \e^{K/2} \e^{t - 4 \alpha} \cos\xi &= 0\,.
\end{split}
\ee
These equations can be used to solve for e.g. the metric function $A$, the scalar $t$, and the phase angle $\xi$.
It is often convenient to write squared BPS equations for the supergravity fields for which the phase $\xi$ drops out
\be\label{squaredEquations}
\begin{split}
(A')^2+\kappa^2 \e^{-2A} - \f{1}9 \e^{K}|\mathcal{W}|^2 &= 0\,,\\
4K_{i\bar\jmath}(z^i)'(\bar z^{\bar\jmath})' -\e^{K}K^{i\bar\jmath}D_i\mathcal{W}D_{\bar\jmath}\overline{\mathcal{W}} &=0\,,\\
(t')^2 - \e^{K}|\partial_t \mathcal{W}|^2 &=0\,,\\
(12\alpha')^2  -\e^{K}|\partial_\alpha \mathcal{W}|^2&=0\,.
\end{split}
\ee

It is worth noting here that the system at hand consists of eight unknown functions, six scalar fields, the metric function $A$, and the phase angle $\xi$. However these are subject to three algebraic equations \eqref{BPS6processed} which reduces the system to five unknown functions. Furthermore, the integral of motion ${\cal J}$ in \eqref{conservedquantitydef} means that at most we must solve four ordinary differential equations. Finally, since the dilaton can be separated out of the system and solved by quadrature, only a system of three non-trivial ODEs must be solved before the dilaton can be determined via a simple integration. The space of solutions is characterized by five integration constants, but only three of those are physical. One can be eliminated by shifting the radial variable $r$ and another can be eliminated by using the scaling symmetry \eqref{scalingsymmetry}.

\subsection{A hidden integral of motion}
We now turn to the full solution to the six-scalar system. To this end we first switch coordinates such that
\be\label{sdef}
\dd s = \f{g}{2}\e^{-4\alpha}\dd r\,.
\ee
Next we note that a particular combination of the superpotential and its derivatives is quite compact, namely
\be
\f13 \mathcal W - \f16 \partial_\alpha \mathcal W -\partial_t \mathcal W = g\e^{-4\alpha-t}z_1z_3\,,\qquad 
\f13 \mathcal W - \f16 \partial_\alpha \mathcal W +\partial_t \mathcal W = -g\e^{-4\alpha+t}\,.
\ee
Inspired by this and using the BPS equations for $A$, $\alpha$, and $t$, we find
\be\label{first-s-der}
\partial_s(A+2\alpha-t) = -\f{\e^{t}\sin\xi}{\sqrt{\Im z_1}\sqrt{\Im z_3}}\,,\qquad \partial_s(A+2\alpha+t) = \f{\Im (\e^{i\xi} z_1 z_3)}{\e^{t}\sqrt{\Im z_1}\sqrt{\Im z_3}}\,,
\ee
where we explicitly wrote out  the exponential of the K{\"a}hler potential.
We see a nice simplification of the right hand side, but the second $s$-derivative brings a further remarkable simplification. This inspires two new variables that satisfy a simple second-order equation
\be\label{secondordereqsforell}
\boxed{~\ell_{\pm} \equiv \f{g}{2}\e^{A+2\alpha\pm t}\,,\qquad \partial_s^2 \ell_\pm = \ell_\pm\,.~}
\ee
We could solve for $\ell_\pm$, but it is useful to pause for a moment here and reflect on what these equations imply. 

Since the two functions $\ell_\pm$ satisfy the same second order linear differential equation with a coefficient of $\ell_\pm'$ that is zero, it follows that the Wronskian $(\partial_s\ell_+) \ell_--\ell_+(\partial_s\ell_-)$ is constant. This follows from the second order equations \eqref{secondordereqsforell} and by itself does not imply anything new since exactly this combination is proportional to the conserved quantity $\mathcal J$ (cf. \eqref{Jresult2}):
\be\label{Wronskian}
(\partial_s\ell_+) \ell_--\ell_+(\partial_s\ell_-) = \tfrac{g^2}2\e^{2A+4\alpha}\partial_s t =  \f{\mathcal J}{\kappa^2}\,.
\ee
In addition to the Wronskian, however, the second order equations \eqref{secondordereqsforell} also imply that 
\be\label{Janus-secondconst}
\kappa^2\big(1-\mu^2\big) \equiv \ell_+\ell_--(\partial_s\ell_-)(\partial_s\ell_+)\,,
\ee
is constant. In this equation we have defined the constant quantity $\mu$. Later on, when we perform a UV expansion of our solution, we will see that this can be identified with the field theory mass.

There are two ways to view the integral of motion \eqref{Janus-secondconst}, both of which will be useful. First, using the algebraic relations \eqref{BPS6processed}, we can express
\be
\ell_+\ell_-=\f{g^2}{4}\e^{2A+4\alpha}= \f{4\kappa^2(\Im z_1) (\Im z_3) }{\Im(\e^{i\xi}(z_1+z_3))^2}\,.
\ee
The right-most term in \eqref{Janus-secondconst} can be simplified using \eqref{first-s-der}
\be
(\partial_s\ell_-)(\partial_s\ell_+) = -\f{4\kappa^2 \sin\xi\,\Im (\e^{i\xi} z_1 z_3)}{\Im(\e^{i\xi}(z_1+z_3))^2}\,.
\ee
Combining these two into the right-hand-side of \eqref{Janus-secondconst} we find 
\be
1-\mu^2 = \f{4\,\Im(\e^{i\xi}z_1)\Im(\e^{i\xi}z_3)}{{\Im(\e^{i\xi}(z_1+z_3))^2}}=1-\f{\Im(\e^{i\xi}(z_1-z_3))^2}{{\Im(\e^{i\xi}(z_1+z_3))^2}}\,.
\ee
In other words, by choosing a branch for the square root, the new integral of motion can be expressed as
\be\label{muzfinal}
\boxed{~\mu = \f{\Im(\e^{i\xi}(z_1-z_3))}{{\Im(\e^{i\xi}(z_1+z_3))}}\,.~}
\ee

\begin{figure}[t]
\centering
\includegraphics[width=0.6\textwidth]{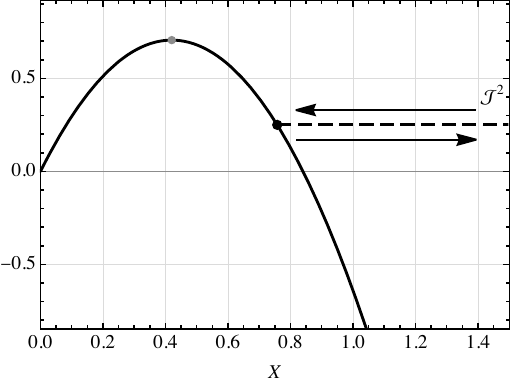}
\caption{\label{fig:potential}The effective potential $V_\text{eff}=4X\big(1-\mu^2-X\big)$ for AdS$_4$ slices, $\kappa^2=1$, at $\mu=0.4$ (solid), together with the energy $\mathcal J^2$  with $\mathcal J=0.5$ (dashed). The particle comes in from large $X$, one asymptotically AdS$_5$ region, bounces off the potential at the turning point (black dot), which is the throat of minimal $X$, and returns to large $X$, the second AdS$_5$ region. The arrows indicate the two legs of this motion. The grey dot is the maximum of the potential, $X=\tfrac12(1-\mu^2)$, $V_\text{eff}=(1-\mu^2)^2$. A particle at rest there would be an AdS$_4\times\R$ solution. Such solutions are  regular only for $\mu=0$, where they are the J-folds of Section~\ref{sec:jfolds}.}
\end{figure}

Now we consider a second viewpoint of eq. \eqref{Janus-secondconst} and define
\be\label{Xdef}
X \equiv \ell_+\ell_-=\f{g^2}{4}\e^{2A+4\alpha}
\ee
and combine \eqref{Janus-secondconst} with \eqref{Wronskian} to write a single ODE for $X$:
\be\label{Particleq}
(\partial_s X)^2+4X\big(\kappa^2(1-\mu^2)-X\big)=\mathcal{J}^2\,.
\ee
This equation can be interpreted as describing a motion of a classical particle in the inverted harmonic oscillator potential $V_\text{eff}=4X\big(\kappa^2(1-\mu^2)-X\big)$ with total energy $E_\text{eff} = \mathcal{J}^2$. Roughly speaking, the particle comes in from asymptotic infinity $X\to\infty$ and approaches the potential wall close to the origin (see Figure~\ref{fig:potential}). Depending on the value of the parameters $\kappa$, $\mu$, and $\mathcal J$, the particle either bounces off the potential wall, approaches a standstill at a critical point of the potential, or crashes down towards $X\to0$ where the system breaks down. All of these cases have a clear physical interpretation in the five-dimensional geometry which we will discuss in more detail below. In short, the bouncing solutions correspond to Janus configurations which have two asymptotic AdS$_5$ regions and a regular turning point in the middle. A very similar structure was found when the original Janus solution was constructed in \cite{Bak:2003jk}. Later, supersymmetric Janus solutions were also shown to be controlled by a similar one-dimensional classical mechanics problem \cite{Bobev:2020fon}. We will show in Section~\ref{sec:janus} that our system reduces in special limits to two of the Janus solutions presented in \cite{Bobev:2020fon}. The solutions for which $X\to0$ are almost always singular in five dimensions as the metric function $\e^{2A}$ approaches zero. Crucially there are regular solutions for which $X$ reaches zero when $X=0$ happens also to be a turning point of the potential. Euclidean configurations on $S^4$ are exactly of this type which we analyze in more detail in Section~\ref{sec:sphere}. Finally, we can have fine-tuned situations where $X$ reaches the potential at a critical point, $X$ will stop there and stay for as long as we like. These are J-fold solutions. 

Even though the hidden integral of motion has already greatly simplified the task of finding a solution to the BPS equations, we are still one ODE short from a full solution. Fortunately there exists a combination of variables that satisfies a simplified BPS equation which is readily solvable. Consider the combination 
\be\label{Ydef}
Y=\f{\kappa g^3}{8}(\sigma_1-\sigma_3)\,,
\ee
where $\sigma_i$ is defined in \eqref{sigmadef}. Now use \eqref{sigmaBPS} to write
\be\label{sigmaBPS1m3}
Y' = \f{\kappa g^3}{4}\e^{3A}\e^{K/2}\Re\Big[\e^{i\xi}( V_1-V_3)\mathcal{W}\Big]= -\f{\kappa g^4}{4}\e^{3A+2\alpha}\e^{K/2}\Im\Big[\e^{i\xi}( z_1-z_3)\Big]\,.
\ee
We can rewrite the right-hand-side using \eqref{muzfinal}, \eqref{BPS6processed}, \eqref{sdef}, and \eqref{Xdef}
\be\label{YBPS}
\boxed{~\partial_s Y =  -4\kappa^2 \mu X\,.~}
\ee
This equation is the final ingredient that allows us to find a general closed form solution to the full BPS system of equations. 
It is interesting to note that the definition of $Y$ is closely related to the definition of $\mathcal{J}$ in terms of scalar fields. If the superpotential were invariant under independent scalings of $z_1$ and $z_3$ (and not just the simultaneous scaling) then $Y$ as defined would also be a conserved charge. A very similar strategy to identify suitable variables was taken in \cite{Astesiano:2023iql} for a closely related but different supergravity model that describes the holographic dual to the equal mass ${\cal N}=1^*$ theory. Indeed it can be shown that in the ten-scalar model, three such $Y$'s can be defined which satisfy simple BPS equations. We briefly discuss this in Section~\ref{sec:discussion}.

\subsection{The solution}
The path to solving the BPS equations is now straightforward. Using \eqref{secondordereqsforell}, we write
\be
\ell_\pm = \f{\e^{\pm t_0}\sqrt{B}}{2}\big(\e^{s-s_0\pm \delta}+\kappa^2\e^{-s+s_0\mp \delta}\big)\,.
\ee
The parameter $t_0$ can be eliminated using the scaling symmetry of the theory, similarly $s_0$ can be absorbed into a shift in the coordinate $s$ and hence we are left with
\be\label{ellsol}
\ell_\pm = \f{\sqrt{B}}{2}\big(\e^{s\pm \delta} + \kappa^2 \e^{-s\mp \delta}\big)\,.
\ee
Notice that we have parametrized the solution such that the two exponentials reduce to $\cosh$ and $\sinh$ for $\kappa^2=1$ and $\kappa^2=-1$ respectively. This will turn out to be useful when we study different classes of solutions below. Recall that we always have $\kappa^4=1$ which we use when we simplify expressions that follow.
The two integrals of motion are now used to fix the parameters $B$ and $\delta$ to
\be\label{Banddelta}
B=\sqrt{(1-\mu^2)^2-\mathcal J^2}\,,\qquad \tanh2\delta=\frac{\mathcal J}{(1-\mu^2)}\,.
\ee
Inserting this into \eqref{Xdef}, we find
\be\label{Xsol}
X = \f{1}{2}\Big(\kappa^2(1- \mu^2)+B\cosh 2s  \Big)\,.
\ee
Now \eqref{YBPS} can be straightforwardly integrated to give
\be\label{Ysol}
Y = -\mu\Big( 2(1-\mu^2)(s-\gamma)+\kappa^2B\sinh 2s  \Big)\,,
\ee
where $\gamma$ is a new integration constant. Note that for $\mu=0$, the field $Y$ is any constant and not necessarily $Y=0$ as the naive limit of \eqref{Ysol} would imply. 

At this stage all differential equations have been solved and the remaining task is to write the solution in terms of the original fields. After some algebra we can combine the BPS equations \eqref{BPS6processed} with the definition of $\mathcal{J}$ \eqref{Jresult1} and $X$ \eqref{Xdef} as well as $Y$ \eqref{Ydef} to derive a closed form result for the real scalars $t$ and $\alpha$
\be\label{talphasol}
\e^{2t} = \f{\ell_+}{\ell_-}\,,\qquad 
\e^{12\alpha}=\f{4X^3}{\Delta^{2}}\,,\qquad \Delta^{2}=\frac{X(4X^2+\mathcal{J}^2-Y^2)}{1-\mu^2}\,.
\ee
With this we can express the five-dimensional metric as
\be
\dd s_5^2 = L^2\bigg(\f{4}{\Delta^{2}}\bigg)^{2/3}\bigg(X^2\dd s^2 + \frac{\Delta^{2}}{4}\dd s_4^2\bigg)\,,
\ee
where $L=2/g$. 
Similarly we can express the original complex scalars as
\be\label{complexscalars}
z_{1}=\frac{2\,(1+\mu)\,\ell_+\,\big[\kappa^{-1}(\mathcal{J}+ Y)+i\Delta\big]}
{4X^2+(\mathcal{J}+ Y)\big(\mathcal{J}-\kappa^2\partial_s X\big)}\,,\qquad z_{3}=\frac{2\,(1-\mu)\,\ell_+\,\big[\kappa^{-1}(\mathcal{J}- Y)+i\Delta\big]}
{4X^2+(\mathcal{J}- Y)\big(\mathcal{J}-\kappa^2\partial_s X\big)}\,.
\ee
We note here that $\mu\mapsto -\mu$ induces $Y\mapsto -Y$, and hence $z_1\leftrightarrow z_3$ which is a discrete symmetry of the six-scalar model. 

Using the expressions for the complex scalars we can also determine the expressions for the physical fields \eqref{physicalscalars}. In particular we find
\be\label{phisol}
\tan(4\phi_3)=\frac{2\kappa\Delta\mathcal{J}}{\kappa^2\Delta^2+Y^2-\mathcal{J}^2} \,,\qquad
\tan(4\phi_1)=\frac{2\kappa\Delta Y}{\kappa^2\Delta^2+\mathcal{J}^2-Y^2} \,.
\ee
It is noteworthy that the ratio of these is particularly simple
\be
\frac{\sin(4\phi_1)}{\sin(4\phi_3)}=\frac{Y}{\mathcal{J}}\,,
\ee
which will be useful later. Finally we can work out the rather complicated expressions for the two remaining scalars $\alpha_1$ and $\alpha_3$
\be
\begin{split}
\e^{8\alpha_1}&=\Big(\frac{1+\mu}{1-\mu}\Big)^2\,
\frac{2X+\kappa^2(\mathcal J+Y)\,\partial_s\log\ell_+}{2X+\kappa^2(\mathcal J-Y)\,\partial_s\log\ell_+}\,
\frac{2X-\kappa^2(\mathcal J-Y)\,\partial_s\log\ell_-}{2X-\kappa^2(\mathcal J+Y)\,\partial_s\log\ell_-}\,,\\ 
\e^{8\alpha_3}&=\frac{4X^2+\kappa^2\Delta^{2}\big(\partial_s\log\ell_+\big)^2}{4X^2+\kappa^2\Delta^{2}\big(\partial_s\log\ell_-\big)^2}\,.
\end{split}
\ee

\subsection{UV expansion}\label{sec:UV}
The foregoing solution was found without any physical input for the integration constants $\mathcal{J}$, $\mu$ and $\gamma$. In order to give them a meaning in the dual field theory, we expand the solution near the AdS boundary and read off the sources and vacuum expectation values. We find that unsurprisingly this matches the complementary analysis where a solution is found order by order by expanding around the boundary.

Let us start by expanding a general solutions of the BPS equations near the UV, in the variable $\epsilon = \e^{-gr/2}$ which is small there, and write the result in terms of the physical scalar fields. We find
\be\label{UVexpansion}
\begin{split}
\phi_1 &= -\mu\,\kappa\epsilon+\mathcal{O}\Big(\epsilon^3\log\epsilon\Big)\,,\\
\phi_3 &= w\,\kappa^3\epsilon^3+\mathcal{O}\Big(\epsilon^5\log\epsilon\Big)\,,\\
\alpha_1 &= \big(-v+2\mu\log\epsilon\big)\kappa^2\epsilon^2 +\mathcal{O}\Big(\epsilon^4\log\epsilon\Big)\,,\\
\alpha_3 &= \tfrac{w}{2}\,\kappa^2\epsilon^2+\mathcal{O}\Big(\epsilon^4\log\epsilon\Big)\,,\\
\varphi &= \varphi_0-\tfrac12\big(3+\mu^2\big)w\,\kappa^4\epsilon^4+\mathcal{O}\Big(\epsilon^6\log\epsilon\Big)\,,\\
\alpha &= \tfrac13\Big(\mu(\mu+v)-2\mu^2\log\epsilon\Big)\kappa^2\epsilon^2
+\mathcal{O}\Big(\epsilon^4(\log\epsilon)^2\Big)\,.
\end{split}
\ee
together with
\be\label{UVwarp}
g^2\epsilon^2\e^{2A}=1+\tfrac23\big(3-\mu^2\big)\kappa^2\epsilon^2+\mathcal{O}(\epsilon^4\log^2\epsilon)\,.
\ee
Note that every power of $\epsilon$ is accompanied by the same power of $\kappa$. At $\kappa=i$ the expansion \eqref{UVexpansion} is the six-scalar truncation of the expansion of the ten-scalar model in \cite{Bobev:2016nua}. For $\kappa=1$ it describes the approach to the $s\to+\infty$ boundary of a Janus solution. The four constants in \eqref{UVexpansion} have a direct field-theory meaning. The leading coefficient of the pseudoscalar $\phi_1$ is the source $\mu$ of the fermion bilinear, i.e. the hypermultiplet mass. Notice that we have judiciously called one of the integration constants in our solution $\mu$. Formally this is distinct from the UV boundary parameter $\mu$ in eq. \eqref{UVexpansion}, but by expanding our solution in the UV, we will find that these are indeed the same. The coefficient $v$ in $\alpha_1$ is the expectation value of the dual dimension-two operator. The coefficient $w$ in $\phi_3$ is the expectation value of the dimension-three operator dual to $\phi_3=-\phi_4$, which has no source. Finally, $\varphi_0$ is the boundary value of the dilaton, the Yang--Mills coupling.

In order to map the boundary data in \eqref{UVexpansion} to the integration constants in our solution, we similarly perform a UV expansion of the analytic solution. To this end we integrate $\dd \log\epsilon/\dd s = -\e^{4\alpha}$ term-by-term at large $s$ fixing the integration constant by the leading order term in \eqref{UVwarp}. The resulting relation between $\log\epsilon$ and $s$ can be inverted, again term-by-term, at large $s$ or small $\epsilon$ resulting in  
\be 
\e^{-s}=\sqrt{B}\epsilon\Big[1+\kappa^2\epsilon^2\Big(\tfrac43\mu^2\log\epsilon+\tfrac23\mu^2(2\gamma+\log B)\Big)
+\mathcal{O}\Big({\epsilon^4\log^2\epsilon}\Big)\Big] \,.
\ee
Plugging this into the expression for $X$, $Y$, and $\Delta$ and from there into the physical scalar fields we recover exactly the expansion \eqref{UVexpansion} with $\mu$ the integration constant identified with the hypermultiplet mass and
\be\label{UVdictionary}
\mathcal{J}=\tfrac12 w\,,\qquad v=-\mu\big(2+2\gamma+\log B\big)\,,\qquad \varphi_0=\delta\,,
\ee
where $B$ and $\delta$ are defined in \eqref{Banddelta}. We will come back to this identification when we consider different classes of solutions below.

\section{Janus solutions}
\label{sec:janus}
We now turn to domain walls with AdS$_4$ slices, that is $\kappa^2=1$. The solutions we will be interested in are so-called Janus solutions which have two asymptotically AdS$_5$ boundaries $s\to\pm\infty$ and are regular in between. Such solutions describe two copies of the dual gauge theory on two half-spaces, glued together along a three-dimensional interface along which the gauge coupling jumps \cite{Bak:2003jk}.  Regularity of the Janus solution imposes a bound on the conserved charge $\mathcal{J}$ which in turn controls the magnitude of the difference of the two asymptotic values of the dilaton. This then translates to a bound on the jump of the Yang--Mills coupling across the interface in the dual field theory. Saturating the bound (at $\mu=0$) leads to so-called J-fold solutions for which all scalar fields, and the metric, are constant while the dilaton still runs. By compactifying such solutions introducing an $\SL(2,\Z)$ duality twist we obtain AdS$_4$ solutions that have their own 3D SCFT dual \cite{Assel:2018vtq}.

\subsection{Regularity and the dilaton jump}\label{sec:janus-reg}
At an asymptotically AdS$_5$ end the warp factor diverges while all scalars approach their constant vacuum values. The two functions $\ell_\pm=\f g2\e^{A+2\alpha\pm t}$ are therefore positive everywhere and grow without bound at both ends. This explains our choice of parametrization in \eqref{ellsol} which for $\kappa^2=1$ reduces to 
\be\label{ellsoljanus}
\ell_\pm = \sqrt{B}\cosh(s\pm \delta)\,.
\ee
For a real solution we furthermore need $B\ge 0$ and $\delta$ real which by the relation \eqref{Banddelta} implies
\be\label{Jregbound}
|\mathcal J|\le 1-\mu^2\,.
\ee
This bound is saturated when $B=0$ and the corresponding solutions which will be discussed separately in Section~\ref{sec:jfolds} below. Here we focus on regular Janus solutions which have $|\mathcal J|<1-\mu^2$ and for real $\mu$ this implies that $|\mu|<1$. The throat of the Janus, where $X$ is the smallest, is located at $s=0$. This is where the classical mechanics particle of Section~\ref{sec:solution} has zero kinetic energy and it bounces off the potential wall.

We can work out the difference of the dilaton values in the two asymptotic regions. Recall that the dilaton is dual to a marginal operator in the dual field theory and hence any vacuum value for it is allowed in the asymptotic region. By \eqref{talphasol} we have $\e^{2t}=\cosh(s+\delta)/\cosh(s-\delta)$, so $t\to\pm\delta$ as $s\to\pm\infty$. Since $\alpha_3$ vanishes at both boundaries, $t=-\alpha_3+\varphi$ shows that the dilaton interpolates between $-\delta$ and $\delta$. This is in agreement with $\varphi_0=\delta$ in \eqref{UVdictionary} which assumed a certain sign for the asymptotic data. The jump is therefore $\Delta\varphi=2\delta$, or by \eqref{Banddelta}
\be\label{jumpformula}
\boxed{~ \tanh\Delta\varphi=\f{\mathcal J}{1-\mu^2}\,.~}
\ee

\subsection{The massless case}\label{sec:janus-massless}
\begin{figure}[t!]
\centering
\includegraphics[width=0.55\textwidth]{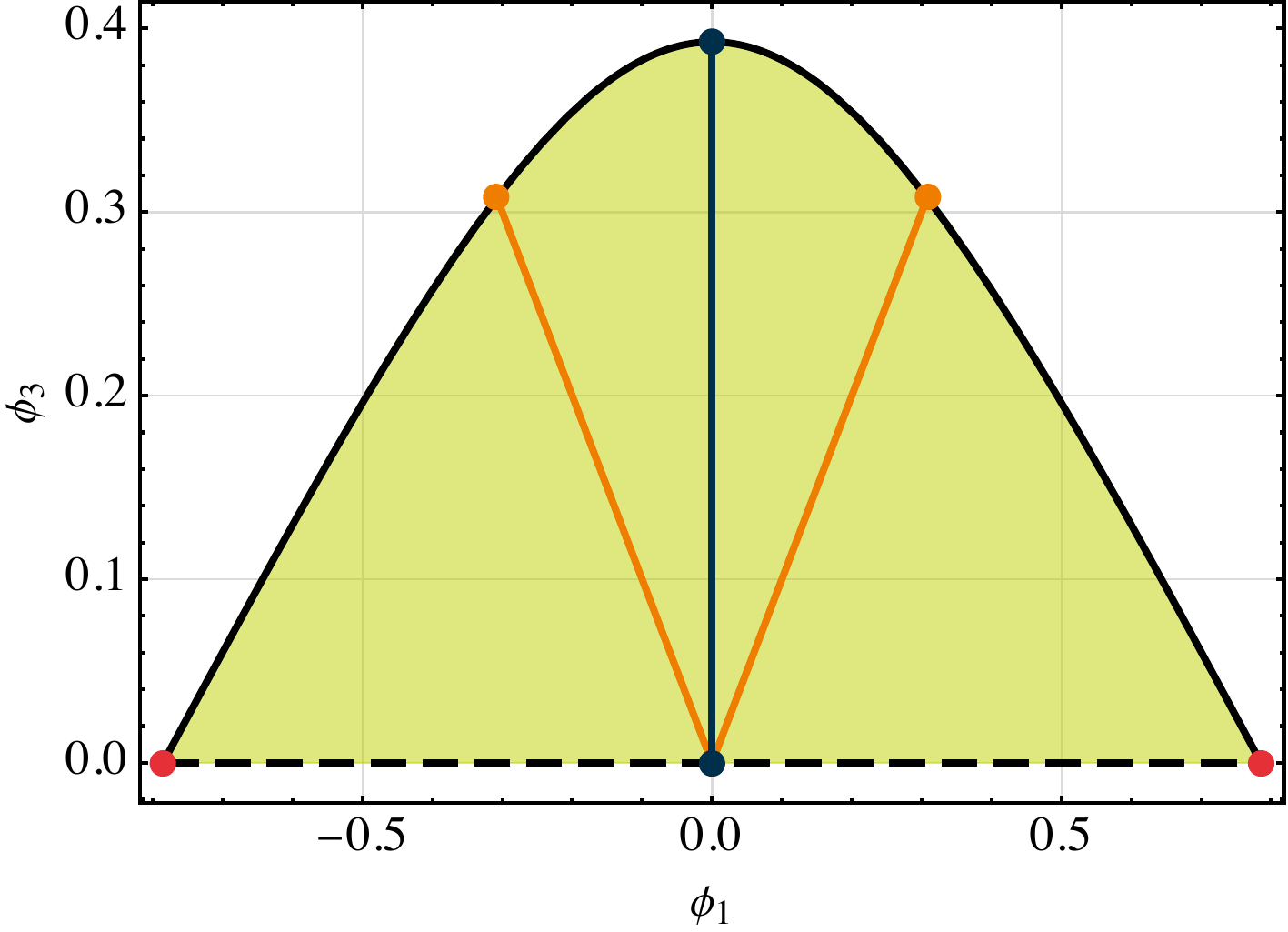}
\caption{\label{turnregion}The region of regular Janus solutions with $\mu=0$ characterized in terms of the throat values of the scalars $\phi_1$ and $\phi_3$. $\phi_1\to-\phi_1$ is a symmetry of the system and physically equivalent solutions are represented twice except on the vertical $\phi_1=0$ axis. The constant of motion $\mathcal{J}$ increases from $\mathcal{J}=0$ on the dashed line up to $\mathcal{J}=1$ on the solid arc at the top of the shaded region. The solid arc also denotes a one-parameter family of J-fold solutions terminating at a singularity at the red dots. The solid vertical line indicates the symmetry enhanced $\SU(2)\times \U(1)$ Janus. The solid orange lines indicate the 3D $\mathcal{N}=4$ invariant Janus with $\SO(3)\times \SO(3)$ symmetry.}
\end{figure}
Let us now focus on the zero mass Janus solutions. By \eqref{YBPS} the function $Y$ then acts as a second constant of motion, and the solution of Section~\ref{sec:solution} becomes\footnote{The reader should be careful not to take $\mu=0$ in \eqref{Ysol} without rescaling $\gamma$ to absorb the $\mu$.}
\be\label{masslessX}
X=\f12\Big(1+\sqrt{1-\mathcal J^2}\cosh2s\Big)\,,\qquad Y=\text{constant}\,,\qquad \tanh2\delta=\mathcal J\,.
\ee
The massless Janus thus form a two-parameter family labelled by $(\mathcal J,Y)$. By \eqref{jumpformula} all solutions with the same $\mathcal{J}$ have the same dilaton jump but are distinguished by $Y$.

Two limits of this family were previously constructed in \cite{Bobev:2020fon}. The first is the $\SU(2)\times\U(1)$ invariant Janus discussed there. Here it can be obtained by setting $z_1=z_3$ which in our parametrization using \eqref{sigmadef} and \eqref{Ydef} is equivalent to $Y=0$ for $\mu=0$. Furthermore the condition  $z_1=z_3$ also ensures that the numerator of \eqref{muzfinal} vanishes, and hence $\mu=0$. Next, the $\SO(3)\times\SO(3)$ invariant, 3D $\mathcal{N}=4$ Janus discussed in \cite{Bobev:2020fon} is obtained in the six-scalar model by setting $z_3=i\e^{t}$ and $\alpha=0$. Here $\Re z_3=\sigma_3=0$ and $Y=\mathcal J$ by \eqref{Jresult1} and \eqref{Ydef}. Since $Y$ is then obviously constant, the only consistent way to solve \eqref{YBPS} is to let $\mu=0$. We note that for the $\mathcal{N}=4$ invariant Janus, we have $\alpha=0$ and therefore $X=\f{g^2}4\e^{2A}$ is just the metric function itself. The classical mechanics equation \eqref{Particleq} for $X$ is then equation (2.17) of \cite{Bobev:2020fon}, where $\mathcal I$ there should be equated with $\mathcal J^2$ here.\footnote{The paper \cite{Bobev:2020fon} also defines $\mathcal{J}_\text{there}$ which does not have a counterpart in this work. When comparing our solutions with \cite{Bobev:2020fon} we must put $\mathcal{J}_\text{there}=0$. This parameter can be reinstated by an $\SL(2,\R)$ transformation as discussed there (see also \cite{Gautason:2020mgd}).} For the $\SU(2)\times\U(1)$ invariant Janus, we may also find a direct link between the classical mechanics equation \eqref{Particleq} and (3.15) in \cite{Bobev:2020fon} after a simple change of coordinates $X_\text{there}=\mathcal J^2/(4X^2_\text{here}+\mathcal J^2)$ and once again $\mathcal I$ there is the same as $\mathcal J^2$ here.

\begin{figure}[t!]
\centering
\includegraphics[width=0.49\textwidth]{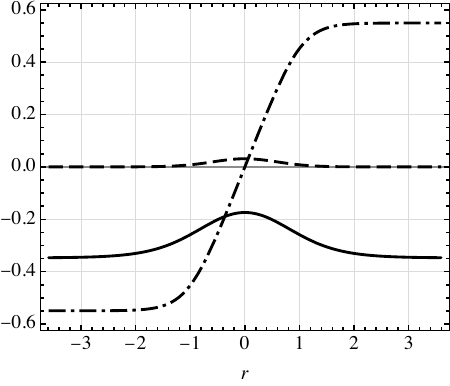}\hfill
\includegraphics[width=0.46\textwidth]{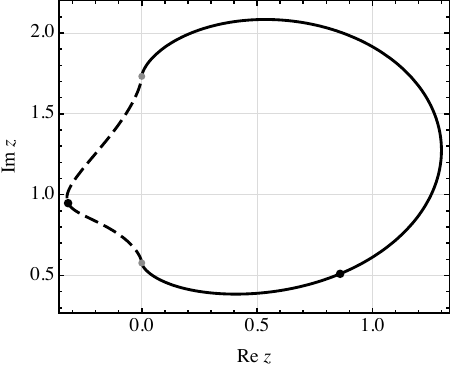}
\caption{\label{fig:janus-massless}A massless Janus wall with $\mathcal J=0.8$ and $Y=1.2$, a member of the family with $Y$ beyond the $\SO(3)\times\SO(3)$ value $Y=\mathcal J$, in units $L=1$. Left: the warp factor measured against the AdS$_4$ slicing of AdS$_5$, $A-\log\cosh r$ (solid), the scalar $\alpha$ (dashed) and the dilaton $\varphi$ (dash-dotted), against the proper distance $r$, with the throat at $r=0$. The dilaton interpolates between $\mp\delta$ with $\tanh2\delta=\mathcal J$, and $\alpha$ is even in $r$, as it is on every massless wall. Right: the trajectories of $z_1$ (solid) and $z_3$ (dashed) in the upper half-plane. The grey dots are the vacuum $z_1=z_3=i\e^{\mp\delta}$ reached at $r\to\mp\infty$, and the black dots mark the throat.}
\end{figure}

These two limiting families constitute only a small subset of the total space of massless Janus solutions. 
Note that $X$ is even in $s$ and $Y$ is constant (for $\mu=0$) and $s\to-s$ exchanges $\ell_+\leftrightarrow\ell_-$, so $t$ is odd. From this it follows that the dilaton $\varphi$, $\alpha_1$ and $\alpha_3$ are odd while the metric $A$ and $\alpha$, $\phi_1$, and $\phi_3$ are even. In particular $\alpha_1=\alpha_3=0$ at the throat, and the massless Janus solutions can be characterized by the value of $\phi_{1,3}$ at the throat instead of the integration constants $\mathcal{J}$ and $Y$ as follows
\be
\mathcal J=\f{2\sin2\phi_3\sqrt{\cos^22\phi_1-\sin^22\phi_3}}{\cos^22\phi_1}\,,
\ee
and $Y/\mathcal J=\sin4\phi_1/\sin4\phi_3$ as seen in Section~\ref{sec:solution}. At the throat $X=1-\sin^22\phi_3/\cos^22\phi_1=\f12(1+\sqrt{1-\mathcal J^2})\ge 1/2$, which implies 
\be\label{jfoldphases}
\cos^2 2\phi_1\ge 2\sin^2 2\phi_3\,.
\ee
This bound is plotted in Figure~\ref{turnregion} and is saturated at the J-folds discussed below. The two known Janus families are plotted as straight lines at $\phi_1=0$ ($Y=0$) and $\phi_1=\phi_3$ ($Y=\mathcal J$). Our analytic solution covers the entire shaded region and a representative member of the new solutions is shown in Figure~\ref{fig:janus-massless}.

\subsection{J-folds}\label{sec:jfolds}
Continuing with the massless case $\mu=0$, we see that the classical mechanics problem \eqref{Particleq} has a constant solution $X=$const when $\mathcal{J}=1$ and $X=1/2$. This corresponds to the saturation of the bound \eqref{jfoldphases} and sits on the upper edge of the plot in Figure~\ref{turnregion}. In this case the warp factor and all scalars except the dilaton are constant. That is the metric is simply AdS$_4 \times \R$ while the dilaton is linear in $r$. By themselves these solutions are not good string theory solutions as the dilaton grows without bound for large $|r|$ which implies that the supergravity approximation breaks down at some point. This issue can be circumvented by compactifying the $\R$ direction employing the $\SL(2,\Z)$ global symmetry of the theory which acts non-trivially on the dilaton to introduce a duality wall for it. Such solutions are often called an S-folds but we refer to them as J-folds due to their close relation to Janus solutions. See \cite{Assel:2018vtq,Bobev:2019jbi,Bobev:2020fon} for more information on the compactification process and \cite{Inverso:2016eet,Assel:2018vtq,Guarino:2019oct,Bobev:2019jbi,Guarino:2020gfe,Bobev:2020fon,Bobev:2021yya,Guarino:2021hrc,Bobev:2021rtg,Arav:2021tpk,Arav:2021gra,Bobev:2023bxs} for further discussion on these and related solutions. 

In our language the solutions are labelled by the constant value of $Y$, which by positivity of $\Delta^2$ at $|\mathcal J|=1$ ranges over $0\le|Y|<\sqrt2$. We can also label the solutions in terms of the constant value of $\phi_1$ using $Y=\sin4\phi_1/\sin4\phi_3$ and $\cos^2 2\phi_1= 2\sin^2 2\phi_3$. The $\SU(2)\times\U(1)$ invariant J-fold is at $Y=\phi_1=0$, where the corresponding Janus family in \cite{Bobev:2020fon} ends. Similarly, the $\mathcal{N}=4$ J-fold is at $Y=1$ or $\cot2\phi_1=\sqrt{2}$. The continuous family of J-folds parametrized by $Y$ was previously found in \cite{Arav:2021gra} in five dimensions and was also constructed in four-dimensional supergravity in \cite{Bobev:2021yya}. In terms of the physical scalars the J-folds have $\alpha_1=\alpha_3=0$, $\e^{6\alpha}=\cot2\phi_3/\sqrt{2}$ and $\varphi = \mathcal{J}s=\pm s$.

We emphasize that the family of J-fold solutions terminates in a singular solution at $Y\to\sqrt2$, where $\phi_1\to\pi/4$ and $\phi_3\to0$ and $\alpha$ diverges.

\begin{figure}[t!]
\centering
\includegraphics[width=0.49\textwidth]{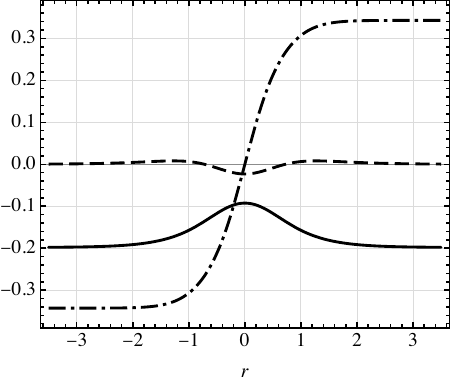}\hfill
\includegraphics[width=0.46\textwidth]{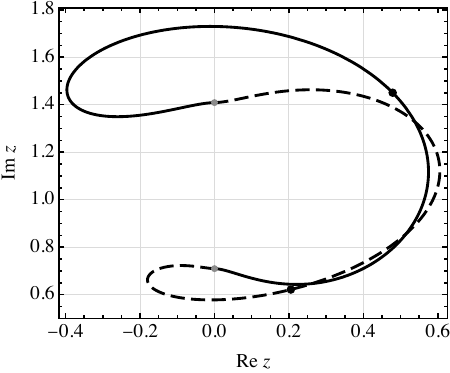}
\caption{\label{fig:janus-massive}A massive Janus interface with $\mathcal J=0.5$, $\mu=0.4$ and $\gamma=0$, in units $L=1$. Left: $A-\log\cosh r$ (solid), $\alpha$ (dashed) and $\varphi$ (dash-dotted) against $r$, with the throat at $r=0$. With $\gamma=0$ $Y$ is odd in $s$, so $\alpha$, which depends on $Y^2$, is even. The asymmetry of the solution is controlled by the pseudoscalars. Right: $z_1$ (solid) and $z_3$ (dashed) in the upper half-plane, with the vacuum at the two ends as grey dots and the Janus throat as black dots. Near each boundary $\Re z_1$ and $\Re z_3$ are proportional to $\phi_1$ and $-\phi_1$, so the two curves reach the vacuum from opposite sides.}
\end{figure} 
\subsection{Janus solutions for \texorpdfstring{$\mathcal{N}=2^*$}{N=2*}}\label{sec:janus-massive}
Finally we consider the Janus solutions with  $\mu\neq0$. These are dual to interfaces where the $\mathcal{N}=2^*$ mass is a function of the coordinate transverse to the interface in such a way that 3D conformal symmetry is maintained \cite{Arav:2020obl}. Regularity requires that not only the inequality \eqref{Jregbound} is satisfied but also that $\Delta^2>0$ along the whole solution. The latter is needed in order for $\alpha$ to remain real. These conditions carve out a complicated allowed region in the space of parameters $(\mathcal J,\mu,\gamma)$ which is difficult to fully specify. Around the massless solutions there is a neighbourhood of massive solutions that are regular. 

We can now compare the two sides of the Janus by performing a UV expansion on each end. In both cases the expansion takes the form \eqref{UVexpansion} but we must be careful to take into account that the orientation of $r$ is opposite on the two sides. This can be fixed by mapping $\kappa\to-\kappa$ at one end before comparing. In this way we see that the two sides of a massive Janus are defined in terms of the same mass $\mu$. More precisely, even though the source of the pseudoscalar $\phi_1$ is $-\mu\kappa$ on the right side $s\to+\infty$ and $+\mu\kappa$ on the left side $s\to-\infty$, these should be interpreted as equal after switching the sign of $\kappa$ on, say, the left side.  Furthermore, the source of the scalar bilinear dual to $\alpha_1$ is the same on both sides. On the other hand the vev of the dimension-two operator $v$ \emph{does} differ between the two sides by $v_+-v_-=-4\mu\gamma$. Figure~\ref{fig:janus-massive} shows this on a representative Janus solution.

Massive Janus solutions with $\mathcal J=0$ were constructed numerically in \cite{Arav:2020obl}, within the three-scalar truncation of \cite{Bobev:2013cja}. At $\mathcal J=0$ the dilaton is constant, and the six-scalar model reduces to the three-scalar truncation of \cite{Bobev:2013cja} (see Section~\ref{sec:sphere}). The numerical solutions of \cite{Arav:2020obl} are therefore the two-parameter subfamily of our solution labelled by $\mu$ and $\gamma$. We note that the region of regular Janus solutions in that case can be specified completely as $|\gamma|<1/|\mu|-\mathrm{arctanh}|\mu|$. Solutions singled out in \cite{Arav:2020obl} in which the vev of the dimension-two operator is the same on both sides are given by our $\gamma=0$.

We note in closing that there are no massive J-folds. The classical mechanics problem \eqref{Particleq} suggests that tuning $|\mathcal J|=1-\mu^2$ leads to a critical point of the effective potential and we can therefore find a constant solution $X=(1-\mu^2)/2$. However now $Y$ is non-trivial since \eqref{YBPS} can be solved by
\be
Y= -2\mu(1-\mu^2) s + \text{const\,.}
\ee
So not only is the dilaton linear, but also $Y$. In particular \eqref{talphasol} implies that eventually $\Delta^2$ will vanish which means that $\alpha$ has a singularity. Indeed, the metric also becomes singular and these solutions are not physical.

\subsection{One-point function of the Lagrangian}
\label{sec:charge}
To close this section we relate the supersymmetric Janus solutions found in this work to the one-point function of the 4D Yang--Mills Lagrangian $\mathcal O_{\mathcal L}$. This operator is the level-four superdescendant of the superconformal primary operator $\mathcal{O}_2$ of dimension 2 of the stress-tensor multiplet. It is exactly marginal and is holographically dual to the ten-dimensional dilaton $\Phi$. We use the conventions of \cite{He:2026qnv} for the normalization of the operator. The functional form of the one-point function of an operator $\mathcal{O}$ of dimension $\Delta_\mathcal{O}$ is fixed by the 3D conformal symmetry preserved by the interface to be $C_{\mathcal{O}}/|x_\perp|^{\Delta_\mathcal{O}}$ and is controlled only by a single number $C_{\mathcal O}$. For the operator $\mathcal O_{\mathcal L}$, this number was computed in QFT using perturbation theory in \cite{Clark:2004sb,Karch:2026ymg} and then \cite{He:2026qnv} showed by supersymmetric localization that for the $\mathcal{N}=4$ and $\mathcal{N}=2$ interfaces the answer is exact in $N$ and in the 't Hooft coupling. The result was shown to match holographic computations for the solutions of \cite{DHoker:2007zhm,Bobev:2020fon} in \cite{Karch:2026ymg,He:2026qnv} and we will extend their result to include all massless Janus solutions discussed above.

Let us review the holographic argument here. The source of $\mathcal O_{\mathcal L}$ is the boundary value of the dilaton, and its expectation value is the momentum conjugate to it. The scaling symmetry \eqref{scalingsymmetry} acts on the physical scalars \eqref{physicalscalars} as $\varphi\to\varphi+c$ with everything else fixed. The momentum conjugate to $\varphi$ is therefore the Noether current, which  from the supergravity action is calculated to be $\partial\mathcal{L}/\partial(\varphi')=-2\e^{4A}J_r$ with $J_r$ the radial component of the current $J$ in \eqref{Jformandconservation}. Hence we should expect that $\langle\mathcal O_{\mathcal L}\rangle$ is proportional to the conserved quantity $\mathcal{J}$. Indeed, the expectation value can be read off from the UV expansion of the ten-dimensional dilaton 
\be\label{10ddilatonexpansiongeneral}
\e^\Phi = \e^{\Phi_\pm}\bigg[ 1+\f{2\pi^2}{N^2}\langle\mathcal{O}_{\mathcal{L}}\rangle f^4 + \cdots\bigg]\,,\qquad f = z\e^{-A}\,, 
\ee
and $z$ is the Poincar\'e coordinate of the AdS$_4$ slice. This prescription reads the expectation value off the dilaton only when it decouples from other scalars near the boundary. In our model this is the case for $\mu=0$ but for $\mu\ne 0$ the dilaton mixes with $\alpha_3$ at the order of the expectation value, and we must take that into account.

Let us therefore focus on the case $\mu=0$. The expansion \eqref{UVexpansion} then gives $\varphi=\varphi_0-3\mathcal J\epsilon^4+\dots$ with $\e^{A}\simeq1/(2\epsilon)$.\footnote{We set $L=1$ or $g=2$ in this subsection.} The ten-dimensional dilaton $\Phi$ is related to the five-dimensional one by $\Phi=2\varphi$ at the boundary,\footnote{See e.g. Appendix C in \cite{Bobev:2016nua} where this is derived. Note however that in \cite{Bobev:2016nua} $\Phi=-2\varphi$ due to a different choice of conventions. Here we use the conventions of \cite{Bobev:2020fon}.} and hence
\be
\e^\Phi = \e^{2\varphi_0}\bigg[ 1-\f{3\mathcal{J}}{8z^4} f^4 + \cdots \bigg]\,,
\ee
which upon comparing with \eqref{10ddilatonexpansiongeneral} gives
\be\label{Ovev}
\langle \mathcal{O}_{\mathcal{L}}\rangle =-\f{3N^2\mathcal{J} }{16\pi^2 }\f{\text{sgn}(x_\perp)}{|x_\perp|^4}\,,
\ee
where we used $z=|x_\perp|$ as the distance from the interface and $\text{sgn}(x_\perp)=\text{sgn}(s)$. Using the expression for the dilaton jump in \eqref{jumpformula} (with $\mu=0$) and the relation between the 10D dilaton and the 't~Hooft coupling $\lambda = 4\pi N\e^{\Phi}$ on the two sides of the interface \cite{Karch:2026ymg,He:2026qnv}
\be\label{thooftjump}
\f{\lambda_+-\lambda_-}{\lambda_++\lambda_-} = \tanh \f{\Delta\Phi}{2} =\tanh\Delta\varphi= \mathcal J\,,
\ee
where the subscript $\pm$ refers to the 't~Hooft coupling evaluated in the asymptotic region where $s\to\pm\infty$. In \cite{He:2026qnv} $\lambda_+$ is denoted by $\lambda_L$ and $\lambda_-$ by $\lambda_R$. We may write this as
\be\label{COL}
C_{\mathcal O_{\mathcal L}} =-\f{3N^2}{16\pi^2 }\f{\lambda_+-\lambda_-}{\lambda_++\lambda_-}\text{sgn}(x_\perp)\,.
\ee
This agrees with the result for half-BPS and the $\SU(2)\times\U(1)$ invariant interfaces established in \cite{Karch:2026ymg,He:2026qnv}. Our computation shows that this holds for all massless Janus solutions of Section~\ref{sec:janus-massless} and is independent of the parameter $Y$.

We point out that this result could have been obtained without going to ten dimensions but by doing a careful holographic renormalization of the action and operators directly in five dimensions. This was done in \cite{Bobev:2016nua} for the ten-scalar model where in eq. (4.35) it was shown that\footnote{We correct for the different normalization and conventions used for $\mathcal{O}_\varphi = -2\mathcal{O}_{\mathcal L}$ in \cite{Bobev:2016nua}.}
\be
\langle \mathcal{O}_{\mathcal L} \rangle =-\f{3 N^2}{2\pi^2}(w-2\mu_1\mu_2\mu_3) = -\f{3 N^2\mathcal{J}}{\pi^2}\,,
\ee
where in the second step we used that in the full ten-scalar model $2\mathcal{J}=w-2\mu_1\mu_2\mu_3$ as shown in Appendix \ref{App:10and6scalars}. Note that this vev was computed for $S^4$ slices but holds without modification on AdS$_4$. The boundary metric is that of AdS$_4$ of radius $1/2$ and the conformal map to the half-space introduces a factor of $(2|x_\perp|)^{-4}$ for an operator of dimension $4$.
After correcting for this fact and reducing to the six-scalar model by setting $\mu_1=\mu_2=\mu$ and $\mu_3=0$, we recover \eqref{Ovev}. This then also shows that the result \eqref{Ovev} holds even for $\mu\ne0$. As noted above, reading the expectation value off the dilaton alone is not reliable for $\mu\ne0$ due to the mixing of operators which is taken into account by the holographic renormalization. When $\mu\ne0$, the relation between the 't~Hooft coupling jump and $\mathcal{J}$ in \eqref{thooftjump} gets corrected to
\be\label{thooftjumpmassive}
\f{\lambda_+-\lambda_-}{\lambda_++\lambda_-} = \f{\mathcal J}{1-\mu^2}\,,
\ee
and hence for the massive Janus interfaces we have
\be
C_{\mathcal O_{\mathcal L}} =-\f{3N^2(1-\mu^2)}{16\pi^2 }\f{\lambda_+-\lambda_-}{\lambda_++\lambda_-}\text{sgn}(x_\perp)\,.
\ee

\section{Spherical domain walls}\label{sec:sphere}
In this section we turn to four-sphere sliced domain walls with  $\kappa^2=-1$. These solutions, studied numerically in \cite{Bobev:2013cja}, are the holographic duals to $\mathcal{N}=2^*$ theory on $S^4$. This is a particularly attractive example as many observables, including the partition function, can be treated using the powerful tool of supersymmetric localization \cite{Pestun:2007rz}. The solution of Section~\ref{sec:solution} can be ported over to this case without much modification. We must only take care of two things. First is the regularity condition imposed by the shrinking four-sphere. A Euclidean domain wall with four-sphere slices has one boundary and must cap off smoothly in the interior in order not to produce a singularity. This condition introduces a relation between the integration constants. Indeed we will find that regularity demands
\be
\mathcal J = \gamma =0\,.
\ee
This still leaves a one-parameter family of solutions labelled by the mass $\mu$ as could have been expected from the dual field theory.

Next we must relax the standard reality conditions of the fields once we are in Euclidean signature. From our perspective this will be a natural outcome of our analytic solution but we will comment on it below.

\subsection{Regularity}\label{sec:sphere-cap}
As mentioned, for a regular domain wall with $S^4$ slices we expect that the energy associated with the curvature of the sphere forces it towards a collapse as a function of the radial coordinate $r$ (or $s$). We therefore must impose that the sphere shrinks to zero size $\e^{A}\to0$ in a regular fashion. In particular all scalars must approach a constant and have a zero derivative at the point of collapse. This means that $\e^{2t}=\ell_+/\ell_-$ approaches a finite value there. Recall that individually $\ell_\pm$ are dominated by the behaviour of the metric as $\e^{A}\to0$ and hence both $\ell_+$ and $\ell_-$ vanish at the same value of $s$ which forces $\delta=0$. Clearly this means that $\ell_\pm\propto \sinh s$, i.e. they are proportional to each other and thus their Wronskian vanishes. By \eqref{Wronskian}
\be\label{Jforsphere}
\mathcal J=0\,,
\ee
with the direct consequence that by \eqref{UVdictionary}, the vacuum expectation value of the operator dual to $\phi_3$ is switched off $w=0$. The condition \eqref{Jforsphere} also implies that $t$ is constant (see e.g. \eqref{Jresult2}). By the scaling symmetry we may set $t=0$ but the $t$-equation in \eqref{BPS6scalars} then requires $\partial_t\mathcal W=0$, that is $z_3=-1/z_1$. At the level of the six-scalar model, setting 
\be
t=0\,,\qquad z_3=-1/z_1\,,
\ee
gives a consistent truncation of the theory to the three-scalar $\mathcal{N}=2^*$ model of \cite{Bobev:2013cja}. The fact that regular spherical domain walls require a truncation to the three-scalar model was already observed in \cite{Bobev:2016nua}. For simplicity of notation, we let $z_1 = z$ in this section. For this truncation, the K{\"a}hler potential and superpotential reduce to
\be\label{3scalarsuperpot}
K = 2\log\tfrac{|z|}{2\Im z}\,,\qquad \mathcal{W} = -g\mathrm{e}^{-4\alpha}+ig\mathrm{e}^{2\alpha}\Big(z-\tfrac1{z}\Big)\,.
\ee
For completeness we write the BPS equations of the reduced system 
\be\label{3scalarBPS}
\begin{split}
\frac{z'}{\alpha'}&=-3 (z-\bar{z})\frac{ 2 \e^{6 \alpha } (z\bar{z}-1)-i (z+\bar{z})}{2 \e^{6    \alpha } (\bar{z}^2-1)+4 i \bar{z}}\,,\\
\frac{\bar{z}'}{\alpha'}&=3 (z-\bar{z})\frac{ 2 \e^{6 \alpha } (z\bar{z}-1)+i (z+\bar{z})}{2 \e^{6\alpha } (z^2-1)-4 i z}\,,\\
(\alpha')^2 &= -\frac{\e^{-8 \alpha } g^2 (\e^{6 \alpha } (z^2-1)-2 iz) (\e^{6 \alpha } (\bar{z}^2-1)+2 i\bar{z})}{36 (z-\bar{z})^2}\,.
\end{split}
\ee
These equations must be supplemented by the algebraic equation for the metric which can be read off from \eqref{BPS6processed}. Let us use the opportunity to introduce variables which are inspired by the uplift to ten dimensions which was carried out in \cite{Bobev:2018hbq} and is reviewed in Section~\ref{10Dbackground-sec}. These will also turn out to be useful when taking the flat space limit of our solution. We define
\be\label{ACT}
C=\frac{i(1+z\bar z)}{z-\bar z},\qquad T=\frac{z\bar z-1}{z\bar z+1}\,,
\ee
which in terms of physical scalars is $C=\cosh(2\alpha_1)/\cos(2\phi_1)$ and $T=\tanh(2\alpha_1)$.

Note that the equations \eqref{3scalarBPS} are completely equivalent to the first-order BPS equations in \cite{Bobev:2013cja} where they were numerically solved to describe the holographic dual to the ${\cal N}=2^*$ theory on $S^4$.\footnote{Recall that in \cite{Bobev:2013cja} the system was written in disc variables on $\SL(2,\R)/\U(1)$ instead of the upper-half-plane variables we use here. As a result the equations there take a slightly different form than presented here (see Appendix \ref{App:10and6scalars}).} These equations were also considered numerically in \cite{Arav:2020obl} to describe holographic Janus solutions in ${\cal N}=2^*$. The Janus solutions have already been discussed and solved in Section~\ref{sec:janus-massive} and we focus on the analytic treatment of the four-sphere solution in this section.

\subsection{The solution}\label{sec:sphere-sol}
\begin{figure}[t]
\centering
\includegraphics[width=0.49\textwidth]{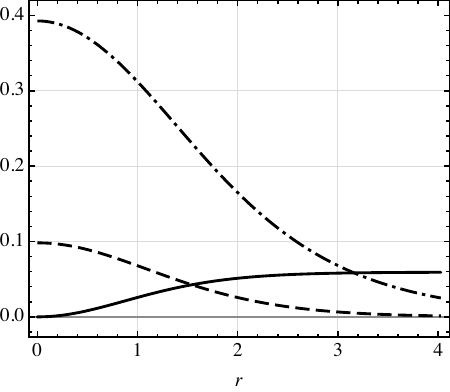}\hfill
\includegraphics[width=0.46\textwidth]{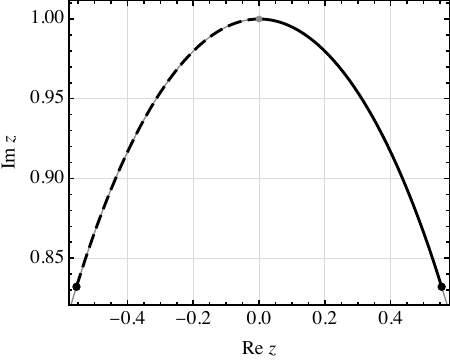}
\caption{\label{fig:sphere}A regular $S^4$ domain wall with $\mu=1.5\,i$ and $\gamma=0$, in units $L=1$. Left: the warp factor measured against Euclidean AdS$_5$ in the $S^4$ slicing, $A-\log\sinh r$ (solid), the scalar $\alpha$ (dashed) and the pseudoscalar $\phi_1$ (dash-dotted), plotted against the radial coordinate $r$. The cap is located at $r=0$. The dilaton is constant and is not shown. Right: $z_1$ (solid) and $z_3=-1/z_1$ (dashed). For imaginary $\mu$ both move on the unit circle (grey), from the vacuum $z_1=z_3=i$ at the boundary (grey dot) to their values at the cap (black dots).}
\end{figure}

Fortunately, we do not have to solve the equations \eqref{3scalarBPS} from scratch, instead we take the solution in \eqref{ellsol}, \eqref{Xsol}, and \eqref{Ysol} and set $\mathcal J=0$ and $\kappa^2=-1$ which yields
\be\label{spheresol}
\begin{split}
\ell_+=\ell_-&=\sqrt{1-\mu^2}\,\sinh s\,,\\
X&=\big(1-\mu^2\big)\sinh^2s\,,\\
Y&=\mu\big(1-\mu^2\big)\big(\sinh2s-2s+2\gamma\big)\,.
\end{split}
\ee
Recall that $Y$ is proportional to $\e^{3A}$ by its definition \eqref{Ydef}, so it must vanish where the sphere shrinks, whereas \eqref{spheresol} gives $Y(0)=2\mu(1-\mu^2)\gamma$. Hence for a regular solution we must have $\gamma=0$. In this case $Y\simeq\f43\mu(1-\mu^2)s^3$ near $s=0$ which vanishes as $\e^{3A}$ should. We will see below that any solution with $\gamma=\mathcal{J}=0$ indeed results in a regular domain wall. An example solution is shown in Figure~\ref{fig:sphere}.

For convenience, let us trade $s$ for $y=\coth s$, so that $\f12\log\f{y+1}{y-1}=s$. In these coordinates we have $X = (1-\mu^2)/(y^2-1)$. Next we define $f(y)$ by writing $Y=2\mu Xf$, the solution for $Y$ in \eqref{spheresol} then reduces to
\be\label{analyticf}
f(y) = y+(1-y^2)\bigg[-\gamma + \f12\log \f{y+1}{y-1}\bigg]\,.
\ee
\begin{figure}[h!]
\centering
\includegraphics[width=0.6\textwidth]{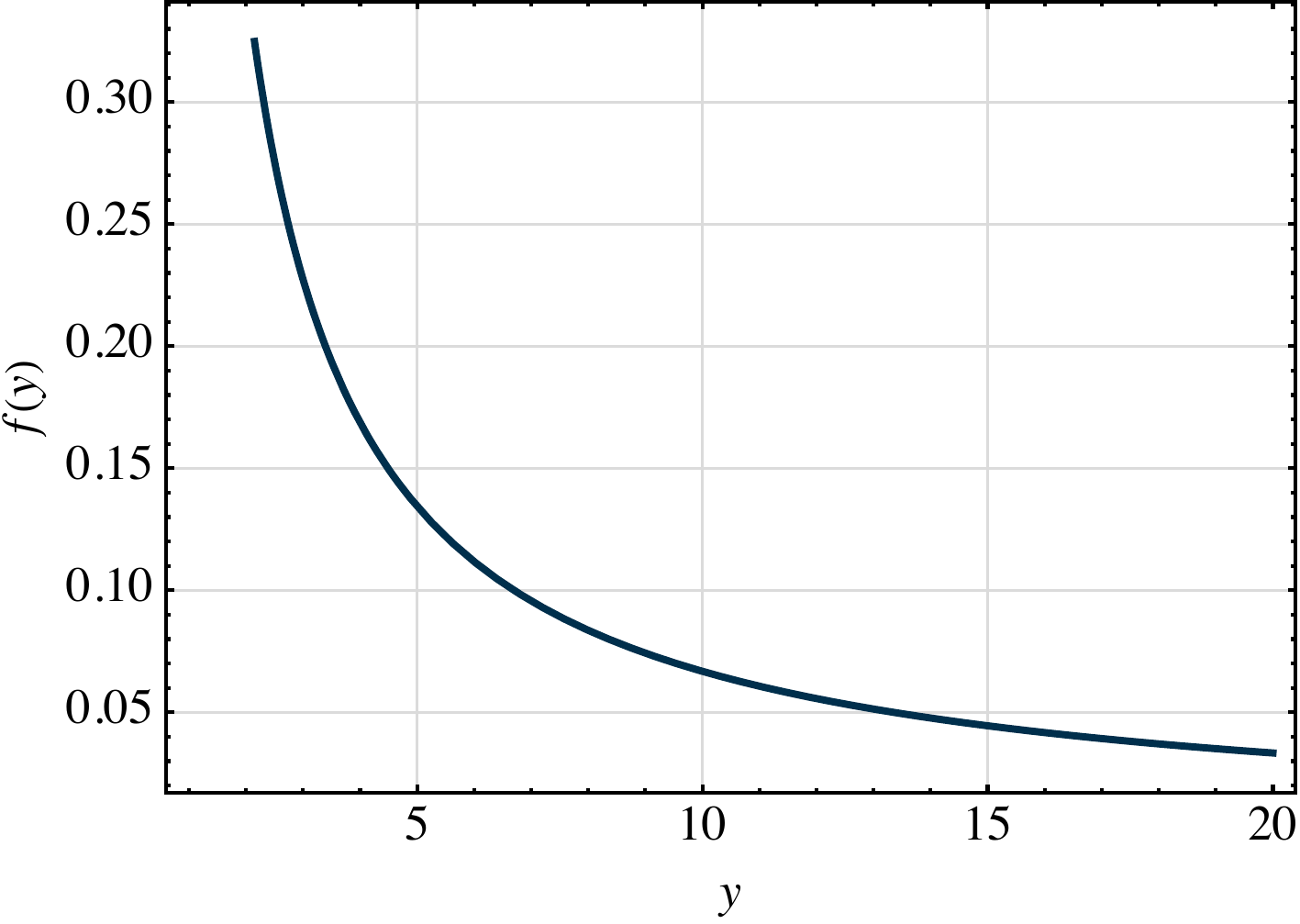}
\caption{\label{fofy}A plot of the function $f(y)$ with $\gamma=0$. The variable $y$ lies in the range $1< y < \infty$.}
\end{figure}
This function $f$ was encountered by Pilch and Warner in \cite{Pilch:2000ue} when they were studying the holographic dual to ${\cal N}=2^*$ on flat space. Here we are working on curved space and although the same function appears, the overall structure of the solution is much more complicated. What this demonstrates however, is that our solutions share some features and our solution approaches theirs in the limit where the four-sphere becomes large. In terms of our variables this is when the dimensionless mass $\mu/i$ is large. As we have explained, the supergravity solution on the sphere is only regular for $\gamma=0$ but we kept $\gamma$ in \eqref{analyticf} for ease of comparison with \cite{Pilch:2000ue}.

Using the change of variables and coordinates above, we may reduce the general expression for $z_1$ in \eqref{complexscalars} to 
\be
\label{zandAintermsof-f-y}
z = \f{i\sqrt{1-\mu^2}}{(\mu-1)(1+y\mu f)}\Big( \sqrt{y^2-1}\mu f-\sqrt{1-\mu^2f^2} \Big)\,.
\ee
Similarly the expression for $\alpha$ in \eqref{talphasol} becomes
\be\label{alphasolfoursphere}
\e^{12\alpha}=\f{1-\mu^2}{1-\mu^2f^2}\,.
\ee
These combined allow us to write $C$ and $T$ as
\be
\label{def-eta-C-T}
C = \f{1-\mu^2 yf}{\sqrt{1-\mu^2}\sqrt{1-\mu^2f^2}}\,,\qquad T = \f{\mu(1-yf)}{1-\mu^2yf}\,,
\ee
and we see that a non-trivial combination of the fields $C$, $T$, and $\alpha$ is related to the integral of motion $\mu$, namely
\be\label{ConstantOfMotion}
\quad\f{C-\e^{6\alpha}}{C T }=\mu\,.
\ee
Finally, for completeness, let us write down the five-dimensional metric in these coordinates
\be
\dd s_5^2 = L^2 \e^{8\alpha}\bigg( \f{\dd y^2}{(y^2-1)^2} + \e^{-12 \alpha}\f{1-\mu^2}{y^2-1}\dd\Omega_4^2 \bigg)\,.
\ee

So far we have not assumed any reality condition on the supergravity fields, and on $S^4$ the usual ones, which hold in Lorentzian signature and for the Janus solutions discussed above, do not hold. With $\kappa=\pm i$ the BPS equations \eqref{BPS6scalars} are complex, and the third algebraic relation in \eqref{BPS6processed} cannot be satisfied by real fields with a real phase $\xi$. This is related to the fact that type IIB supergravity has no real Euclidean form \cite{Hull:1998ym,Hull:1998vg}. In this paper we only demand that the five-dimensional metric must be real. By \eqref{spheresol} and the expression for $\e^{12\alpha}$ above, this holds for real $y\in\R$ with $1\le y < \infty$ exactly when $\mu^2$ is real and bounded above by one.

There are two ways to achieve this. First, if $\mu$ is purely imaginary, $\alpha_1$ is imaginary while $\phi_1$ is real, and $z$ moves on the unit circle. This is the $\mathcal{N}=2^*$ theory on $S^4$ in the usual sense, with $\mu/i$ the \emph{Lorentzian} hypermultiplet mass. If on the other hand $\mu$ is real, $\alpha_1$ is real and $\phi_1$ is imaginary. In Lorentzian language the field theory mass is now tachyonic, but since it is bounded to $|\mu| < 1$, it is allowed in Euclidean QFT. Both of these branches were found numerically in \cite{Bobev:2013cja}, and our analytic solution covers both as well.

Let us analyze the two asymptotic regions. If we let $y=1+2(1-\mu^2)\epsilon^2$ we find that the metric is asymptotically locally AdS$_5$
\be
\dd s_5^2\to L^2 \bigg( \f{\dd y^2}{(y^2-1)^2} + \f{1-\mu^2}{y^2-1}\dd\Omega_4^2 \bigg)\,.
\ee
We can expand our solution for $\alpha$ in \eqref{alphasolfoursphere} to quadratic order in $\epsilon$ and compare with the general expansion in \eqref{UVexpansion}, we find 
\be\label{vofmu}
v = -2\mu - 2\gamma\mu-\mu\log(1-\mu^2)\,.
\ee
The IR regularity condition $\gamma=0$ allows us to recover the relation between the mass $\mu$ and the vev $v$ which was obtained numerically in eq. (4.4) in \cite{Bobev:2013cja} and confirmed perturbatively for small mass in \cite{Kim:2019feb}.\footnote{See also \cite{Kim:2019rwd} for a similar perturbative analysis of other limits of the ten-scalar model dual to the $\mathcal{N}=1^*$ theory.} From this we can also deduce that the on-shell action computed in \cite{Bobev:2013cja} using \eqref{vofmu} can be carried over to our analytic solution as long as the solution is indeed regular. This is easy to check; in the opposite regime $y\to \infty$, we have
\be
f(y) = \f{2}{3y}+ \mathcal{O}(y^{-2})\,,\qquad \e^{12\alpha} = 1-\mu^2 + \mathcal{O}(y^{-2})\,,
\ee
and the metric caps off smoothly as expected
\be
\begin{split}
\dd s_5^2 &\to L^2 (1-\mu^2)^{2/3}\bigg( \f{\dd y^2}{y^4} + \f{\dd\Omega_4^2}{y^2} \bigg)\\
&= L^2 (1-\mu^2)^{2/3}( \dd r^2 + r^2\dd\Omega_4^2 )\,,
\end{split}
\ee
where we momentarily used $r = 1/y$ as a local coordinate around the cap.

As mentioned, the expressions for $C$ and $T$ are useful in order to write down the ten-dimensional uplift of the five-dimensional solution. However they are also useful when studying the large mass limit. Recall that imaginary $\mu$ corresponds to the usual Lorentzian mass and hence we must expand $\mu$ around $i\infty$ to access the large mass limit for which we find
\be\label{PWlimit}
\e^{-6\alpha}\approx f(y)\,,\qquad C \approx y\,,\qquad T \approx 0\,.
\ee
These neatly reproduce the flat space solution of \cite{Pilch:2000ue} where $c_\text{there} = C$ and $\rho_\text{there} = \e^{-\alpha}$.

\subsection{Uplift to type IIB supergravity}
\label{10Dbackground-sec}
We close this section by summarizing the ten-dimensional type IIB solution written down in \cite{Bobev:2018hbq} in terms of the five-dimensional supergravity fields introduced above. We largely follow the notations and conventions of \cite{Bobev:2018hbq}.

The ten-dimensional metric in string frame is
\be\label{metric}
\dd s_{10}^2 = \sqrt{\Xi}\Big(\dd s_5^2 + L^2\dd \widetilde\Omega^2_5\Big)\,,
\ee
where the deformed metric on $S^5$ is written in terms of five frame fields as follows
\be
\dd \tilde\Omega_5^2 = (e^{\hat1})^2 + (e^{\hat2})^2+ (e^{\hat3})^2+ (e^{\hat4})^2+ (e^{\hat5})^2\,,
\ee
and
\be
\begin{split}
e^{\hat1} &= \frac{\e^{\alpha}}{\sqrt{S}}\dd\theta\,,\qquad e^{\hat2} = \frac{\e^{-2\alpha}C}{2\sqrt{SK_1}}\cos\theta\sin\omega\,\big(T\dd \xi_1-\dd\xi_2\big)\,,\\
e^{\hat3} &= \frac{\e^{-2\alpha}}{2\sqrt{SK_1}}\big(-2CT\sin\omega\,\sin\theta\,\dd \theta+S\cos\theta\,\dd \omega\big)\,,\\
e^{\hat4} &= \frac{\e^{-2\alpha}}{2\sqrt{SK_2}} \cos\theta\big(\dd\xi_1+\cos\omega\,\dd\xi_2\big)\,,\qquad
e^{\hat5} = \frac{\e^{\alpha}\sin\theta}{\sqrt{K_2}}\dd \phi \,.\\
\end{split}
\ee
For notational simplicity, we have defined the short-hand functions
\be
\label{def-Cpm}
\begin{split}
\Xi &=\e^{2\alpha}(K_1\sin^2\phi+K_2\cos^2\phi)\,,\\
K_1 &= S\cos^2\theta +  C^2(1-T^2)\e^{-6\alpha}\sin^2\theta\,,\\
K_2 &= S \cos^2\theta +  \e^{-6\alpha}\sin^2\theta\,,\\
S &= C(1+T \cos\omega)\,.
\end{split}
\ee
Note that compared to \cite{Bobev:2018hbq}, we have absorbed one power of $C$ into the definitions of $S$ and $K_1$.

The metric on $S^5$ is highly deformed by the presence of the five-dimensional supergravity fields. The undeformed AdS$_5\times S^5$ solution is obtained by setting $C=1$ and $T=\alpha=0$. This implies $S=K_1=K_2=\Xi = 1$ and the round metric on the five-sphere takes the form
\be
\dd \widetilde\Omega_5^2 \to \dd\theta^2+\f14 \cos^2\theta\,\Big(\dd \omega^2  + \sin^2\omega\,\dd \xi_2^2  +  (\dd \xi_1+\cos\omega\,\dd \xi_2)^2\Big) + \sin^2\theta\, \dd\phi^2\,.
\ee
From this we can read off the coordinate ranges, $0\le \theta \le \pi/2$, $0\le\omega\le \pi$, $0\le \phi,\xi_2 \le 2\pi$, $0\le \xi_1\le 4\pi$. 

In general, the ten-dimensional metric is only $\SO(5)\times\U(1)^2$ invariant. Here $\SO(5)$ is the isometry group of $\dd\Omega_4^2$ and the two $\U(1)$s are generated by $\partial_{\xi_1}$ and $\partial_{\xi_2}$.  It turns out however, that the Einstein-frame metric is invariant under one additional $\U(1)_\phi$ which is generated by $\partial_\phi$. Einstein frame is important for this observation as it is not transformed under the $\SL(2,\R)$ symmetry of type IIB supergravity. The full background is in fact invariant under a diagonal combination of $\U(1)_\phi$ and the compact $\U(1)\subset \SL(2,\R)$ of type IIB supergravity. 

The axion and dilaton are
\be
\label{dilaton-10D}
C_0 = \f{\e^{-4\alpha}(C^2(1-T^2) -1)\sin^2\theta\sin2\phi}{2g_s\Xi}\,,\qquad \e^{2\Phi} = \f{g_s^2\e^{-4\alpha}\Xi^2}{K_1K_2}\,.
\ee
Finally we have the form fields
\be\label{Bfield-and-C}
\begin{split}
B_2 &= L^2\sqrt{C^2(1-T^2)-1}\Big( \sin\phi \,\Psi_1+\cos\phi \,\Psi_2 \Big)\,,\\
g_sC_2 &= L^2\sqrt{C^2(1-T^2)-1}\Big(\cos\phi \,\Psi_1-\sin\phi\, \Psi_2 \Big)\,,\\
C_4^\text{int} &= -\f{L^4\e^{6\alpha}(K_1+K_2)}{2g_s \sin^{2}\theta}\, \Psi_1\w \Psi_2\,.
\end{split}\ee
where we have defined the two 2-forms
\be
\Psi_1 = \sqrt{S}\e^\alpha \cos\theta\, e^{\hat{4}}\w e^{\hat{5}}\,,\qquad \Psi_2 =  \sqrt{K_2}\e^\alpha e^{\hat{1}}\w e^{\hat{4}} + \e^{-2\alpha}\sin\theta\, e^{\hat{2}}\w e^{\hat{3}} \,.
\ee
The form field strengths are defined by
\be\label{field-F}
\begin{split}
H_3&=\dd B_2\,,\quad F_1 = \dd C_0\,,\quad F_3 = \dd C_2 - H_3 C_0\,,\\
{\cal F}_5 &= \dd C_4^\text{int} - \f12(H_3\w C_2 - B_2 \w \dd C_2)\,.
\end{split}
\ee
The five-form ${\cal F}_5$ is not self-dual as constructed and so the full IIB five-form is given by
\be
F_5 = {\cal F}_5 + \star_{10}{\cal F}_5\,.
\ee

Let us take a moment here to discuss the implication of some of the five-dimensional fields being imaginary (or complex). First, let us focus on the mass $\mu$ being real. In this case the five-dimensional fields $C$, $T$, and $\alpha$ are all real and the ten-dimensional metric is also real. Some of the form fields are imaginary but this is expected behaviour for a Euclidean type IIB solution. We can fit our solution into the type IIB$^*$ theory \cite{Hull:1998vg,Hull:1998ym} by analytically continuing the angle $\phi$ which renders the ten-dimensional metric Minkowskian albeit with the time-signature in the `internal' space. This behaviour is consistent with the dual being a deformation of Euclidean ${\cal N}=4$ SYM \cite{Pestun:2007rz}. Indeed, in Euclidean signature, the R-symmetry of ${\cal N}=4$ is $\SO(1,5)$. 

For imaginary $\mu$ the situation is a bit more complicated. Although both $C$ and $\alpha$ are real, $T$ is imaginary. For real angle $\omega$ the ten-dimensional metric therefore appears to be complex. This could be remedied by analytically continuing the angle $\omega = \pi/2 + i \omega_R$ and $\xi_2 = i \xi_{2,R}$ such that $\cos\omega \mapsto -i\sinh \omega_R$. If we furthermore assume $\omega_R,\xi_{2,R} \in \R$, then $S\in \R$ and the metric is real. The price we have to pay is that the frame fields $e^{\hat2}$ and $e^{\hat3}$ are both imaginary, and hence the internal space metric has two time directions.

\section{Discussion}\label{sec:discussion}
In this paper we have found analytic solutions for curved domain walls in the six-scalar model which is designed to study the  $\mathcal{N}=2^*$ theory using holography \cite{Bobev:2016nua}. We have found a three-parameter family of solutions when the domain wall slices are AdS$_4$ corresponding to $\mathcal{N}=2$ Janus interfaces, and a one-parameter family of solutions for spherical slices. All our solutions are regular. In some cases the solutions we present were already studied numerically in the literature and for others, analytic subfamilies were known. We have carried out the UV expansion of our solutions and verified the relation between source and vev found numerically for the spherical domain wall in \cite{Bobev:2013cja}. This is the key input needed to match the scheme-independent part of the free energy to supersymmetric localization. Another comparison that can be carried out is to the 1/2-BPS Wilson loop which similarly to the free energy can be accessed through localization. The classical analysis was already carried out in \cite{Buchel:2013id,Bobev:2018hbq}, but with the analytic solution at hand the one-loop correction is now within reach. The flat limit of this computation was already carried out in \cite{Chen-Lin:2017pay} but we hope to report on the one-loop calculation on the sphere soon \cite{futurework2}.  For the Janus solutions we computed the one-point function of the Lagrangian for both massless as well as massive interfaces. For the massless Janus interfaces these match with the recent works of \cite{He:2026qnv} but it would be interesting to carry out a similar field theory computation for the massive Janus interfaces.

The analytic control of our solution rests on the hidden integral of motion \eqref{muzfinal} which seems to be a feature of the BPS equations rather than a result of a symmetry of the model. The combination \eqref{muzfinal} has a natural extension to the ten-scalar model which is controlled by the superpotential (see Appendix~\ref{App:10and6scalars})
\be
\mathcal{W} =  g\Big(\e^{-4\alpha}(z_1z_3+z_2z_4)+\e^{2\alpha-2\beta}(z_1z_2+z_3z_4)+\e^{2\alpha+2\beta}(z_1z_4+z_2z_3)\Big)\,.
\ee
We may define three functions 
\be
\mu_1 = \f{\Re(\e^{i\xi}(z_1z_2-z_3z_4))}{\Re(\e^{i\xi}(z_1z_2+z_3z_4))}\,,\quad
\mu_2 = \f{\Re(\e^{i\xi}(z_1z_4-z_2z_3))}{\Re(\e^{i\xi}(z_1z_4+z_2z_3))}\,,\quad
\mu_3 = \f{\Re(\e^{i\xi}(z_1z_3-z_2z_4))}{\Re(\e^{i\xi}(z_1z_3+z_2z_4))}\,,
\ee
where $\xi$ is defined in the same way for the domain wall solutions in the ten-scalar model as for the six-scalar model. The names of these functions are not chosen at random, their leading order expansion on the AdS$_5$ boundary is given by the three $\mathcal{N}=1^*$ masses. In the truncation to six scalars, $z_2=z_4=i\e^{t}$, we see that $\mu_1=\mu_2=\mu$ where $\mu$ is defined as in \eqref{muzfinal} and $\mu_3$ vanishes by the first algebraic identity in \eqref{BPS6processed}. Although the ten-scalar BPS equations can be recast in terms of these variables (and three new ones similar to the $Y$ in \eqref{Ydef}), it is clear that $\mu_{1,2,3}$ as defined above are not integrals of motion in the full system except if it is truncated to the six-scalar model considered in this paper. Still, it would be interesting to understand if these new variables illuminate the BPS domain walls of the ten-scalar model and if they can be directly related to the RG flow of the corresponding field theory sources.

\noindent \paragraph{Note added:}~As this work was nearing completion I was informed by Cameron Gibbson that he has independently obtained the analytic solution in Section~\ref{sec:sphere}.

\noindent \paragraph{Acknowledgements:}~I would like to thank Davide Astesiano, Daniel Baig, Nikolay Bobev, and Silviu Pufu for useful discussions. I am especially grateful to Valentina Giangreco M. Puletti and Kostya Zarembo for many inspiring discussions, encouragement and collaboration on related work. I am supported in part by the Science and Technology Facilities Council (Consolidated Grant ST/C004240/1). This work benefited from AI use; I thank Anthropic for providing access to Claude AI. I take full responsibility for the content of the paper.

\appendix
\section{The ten-scalar model and its truncation to six scalars}\label{App:10and6scalars}
In this appendix we briefly review the ten-scalar model of \cite{Bobev:2016nua}. We will use slightly different conventions to parametrize the scalar manifold and provide a map to the original formulation.

As the name suggests, the scalars of the model parametrize a ten-dimensional subspace of the full 42-dimensional scalar manifold $E_{6(6)}/\USp(8)$. This manifold is particularly simple as it contains six commuting factors:
\be
{\cal M} = \OO(1,1)^2\times \Big(\f{\SU(1,1)}{\U(1)}\Big)^4\,.
\ee
The scalars consist of five scalars belonging to the ${\bf 20}'$ representation of $\SO(6)$, four scalars belonging to the ${\bf 10}\oplus\overline{\bf 10}$ and the dilaton (which is a singlet). There are two convenient ways to parametrize the manifold, let us summarize both. Let $\mathfrak{g}_\alpha$, $\mathfrak{g}_\beta$ be the generators of the two $\OO(1,1)$ factors inside $E_{6(6)}$. We then define $\mathfrak{e}_i$, $\mathfrak{f}_i$, and $\mathfrak{h}_i$ as the three generators of $\SU(1,1)$ where $i=1,2,3,4$. They satisfy the standard algebra
\be
[\mathfrak{e}_i,\mathfrak{f}_i]=\mathfrak{h}_i\,,\quad [\mathfrak{h}_i,\mathfrak{f}_i]=-2\mathfrak{f}_i\,,\quad [\mathfrak{h}_i,\mathfrak{e}_i]=2\mathfrak{e}_i\,,
\ee
and as $27\times 27$ matrices embedded inside $\ess$ the Cartan generators are normalized such that $\Tr (\mathfrak{h}_i^2) = 12$. The generators $\mathfrak{e}_i$ and $\mathfrak{f}_i$ are the positive and negative root generators respectively. From them we can define the non-compact and compact generators as follows
\be
\mathfrak{t}_i = \f12(\mathfrak{e}_i+\mathfrak{f}_i)\,,\qquad \mathfrak{r}_i = \f12(\mathfrak{e}_i-\mathfrak{f}_i)\,.
\ee
Explicit construction of the generators just described can be found in section 2 of \cite{Bobev:2020ttg}. 

As mentioned, there are now two convenient ways to parametrize the manifold. The first method was used in \cite{Bobev:2016nua,Bobev:2020ttg} where the $\Ess$ matrix is written as
\be
{\cal V} = \e^{\alpha \mathfrak{g}_\alpha}\cdot \e^{\beta \mathfrak{g}_\beta}\prod_i\e^{-\omega_i \mathfrak{r}_i}\cdot\e^{\rho_i \mathfrak{t}_i}\cdot \e^{\omega_i \mathfrak{r}_i}\,.
\ee 
The scalars $\rho_i$ and $\omega_i$ can be combined into complex fields
\be
\zeta_j = i \tanh\f{\rho_j}{2}\,\e^{-i \omega_j}\,,
\ee
in terms of which the five-dimensional theory studied in \cite{Bobev:2016nua} is defined by a K{\"a}hler potential and a superpotential
\be
\begin{split}
K_\text{disc} &= -\sum_i \log(1-|\zeta_i|^2)\,,\\
\mathcal{W}_\text{disc} &=  \f{g}{2}\Big(\e^{-4\alpha}(1+\zeta_1\zeta_2-\zeta_1\zeta_3-\zeta_1\zeta_4-\zeta_2\zeta_3-\zeta_2\zeta_4+\zeta_3\zeta_4+\zeta_1\zeta_2\zeta_3\zeta_4)\,\\
&\quad +\e^{2\alpha-2\beta}(1-\zeta_1\zeta_2+\zeta_1\zeta_3-\zeta_1\zeta_4-\zeta_2\zeta_3+\zeta_2\zeta_4-\zeta_3\zeta_4+\zeta_1\zeta_2\zeta_3\zeta_4)\,\\
&\quad+\e^{2\alpha+2\beta}(1+\zeta_1\zeta_2+\zeta_1\zeta_3+\zeta_1\zeta_4+\zeta_2\zeta_3+\zeta_2\zeta_4+\zeta_3\zeta_4+\zeta_1\zeta_2\zeta_3\zeta_4)\Big)\,.
\end{split}
\ee
Compared to \cite{Bobev:2016nua}, we have changed names of the scalars in order not to clash with notation of the main text. The scalars $\alpha$ and $\beta$ here are called $\beta_1$ and $\beta_2$ there respectively and the complex fields $\zeta_j$ here are called $z_j$ there.

An alternative parametrization of the manifold, which we use in this paper, is achieved by 
\be
{\cal V} = \e^{\alpha \mathfrak{g}_\alpha}\cdot \e^{\beta \mathfrak{g}_\beta}\prod_i\e^{ \mathfrak{e}_i\Re z_i }
\cdot \e^{\f12 \mathfrak{h}_i\log \Im z_i  }\,.
\ee
Now a different K{\"a}hler potential and superpotential is obtained
\be
\begin{split}
K &= -\sum_i \log(2\Im z_i)\,,\\
\mathcal{W} &=  g\Big(\e^{-4\alpha}(z_1z_3+z_2z_4)+\e^{2\alpha-2\beta}(z_1z_2+z_3z_4)+\e^{2\alpha+2\beta}(z_1z_4+z_2z_3)\Big)\,.
\end{split}
\ee
The explicit map between the coordinates $z_i$ and $\zeta_i$ on the scalar manifold is as follows
\be
\zeta_1=\f{z_1-i}{z_1+i}\,,\quad \zeta_2=-\f{z_2-i}{z_2+i}\,,\quad \zeta_3=-\f{z_3-i}{z_3+i}\,,\quad \zeta_4=\f{z_4-i}{z_4+i}\,,
\ee
under which the superpotential in one set of variable is mapped to a K{\"a}hler transformed version of the superpotential in the other. 

The full scalar-gravity action is
\be
S = \f{1}{16\pi G_N} \int \star\Big(R-12 (\dd\alpha)^2-4(\dd \beta)^2-2 K_{i\bar\jmath}\dd z^i\cdot\dd {\bar z}^{\bar\jmath}-{\cal P}\Big)\,,
\ee
where $K_{i\bar\jmath} = \partial_i\partial_{\bar\jmath}K$ and
\be
{\cal P} =  \f12 \e^{K}\Big(\f16 |\partial_\alpha \mathcal{W}|^2+\f12 |\partial_\beta \mathcal{W}|^2 + K^{i\bar\jmath}D_i\mathcal{W} D_{\bar\jmath}\overline{\mathcal{W}} - \f83 |\mathcal{W}|^2\Big)\,.
\ee

The scalar fields $z_i$ are a mixture of ${\bf 20}'$ and ${\bf 10}\oplus\overline{\bf 10}$ scalars. These in turn are dual to operators in the energy-momentum tensor multiplet of the dual theory and it is therefore convenient to find a direct decomposition of the complex scalars to the physical scalars which have a clear dual in the field theory. This map is obtained via
\be
\begin{split}
\log (-i z_1) &= \alpha_1+\alpha_2+\alpha_3+\varphi - i\phi_1 - i\phi_2 - i\phi_3 + i\phi_4\,,\\
\log (-i z_2) &= -\alpha_1+\alpha_2-\alpha_3+\varphi + i\phi_1 - i\phi_2 + i\phi_3 + i\phi_4\,,\\
\log (-i z_3) &= -\alpha_1-\alpha_2+\alpha_3+\varphi + i\phi_1 + i\phi_2 - i\phi_3 + i\phi_4\,,\\
\log (-i z_4) &= \alpha_1-\alpha_2-\alpha_3+\varphi - i\phi_1 + i\phi_2 + i\phi_3 + i\phi_4\,,
\end{split}
\ee
where we have used the same notation for physical scalars as in \cite{Bobev:2016nua}. In particular the 5D dilaton is denoted by $\varphi$ and it can be checked that it does not appear in the scalar potential. This is a consequence of the $\SL(2,\R)$ symmetry inherited from the ten-dimensional theory. Since the ten-scalar model keeps the dilaton (and not the axion) the $\SL(2,\R)$ symmetry is reduced to a scaling symmetry. Indeed, this is just the scaling symmetry of the six-scalar model discussed in the main body. Here it acts as $\varphi \mapsto \varphi + \log\lambda$ where $\lambda$ is a real constant. We can also write it in terms of the complex fields as
\be
z_i \mapsto \lambda z_i\,,\quad K\mapsto K-4\log(\lambda)\,,\quad K_{i\bar\jmath}\mapsto \lambda^{-2}K_{i\bar\jmath}\,,\quad \mathcal{W}\mapsto \lambda^2 \mathcal{W}\,.
\ee
The symmetry implies a conserved current
\be
J= K_{i\bar\jmath} \dd (z^i{\bar z}^{\bar\jmath})\,,\quad \dd\star J =\f12\star(z^i\partial_i {\cal P}+\bar z^{\bar\jmath}\partial_{\bar\jmath}{\cal P})=0\,.
\ee
It can be shown by extending the UV expansion of \cite{Bobev:2016nua} to higher order that for BPS domain walls, this conserved quantity is related to the three sources $\mu_{1,2,3}$ and the vev $w$ as follows
\be
\mathcal{J}= \f{g^3}{12}\e^{4A}J_r = \f12(w-2\mu_1\mu_2\mu_3)\,,
\ee 
where $\mu_{1,2,3}$ are the sources for the fields $\phi_{1,2,3}$ and $w$ is the vev for $\phi_4$ \cite{Bobev:2016nua}. This reduces to the equation \eqref{UVdictionary} if we set $\mu_1=\mu_2$ and $\mu_3=0$.

The system of supersymmetry variations leads to equations for a Killing spinor and for the fields. These read \cite{Bobev:2016nua}
\be\label{10scalarspinorvar}
\begin{split}
\nabla_\mu \epsilon_1 - \f{i}4 Q_\mu \epsilon_1 - \f{i}6 \e^{K/2}{\overline{\mathcal{W}} }\gamma_\mu \epsilon_2 &= 0\,,\\
(\slashed \partial z^i) \epsilon_1 + \f{i}2 \e^{K/2}K^{i\bar\jmath}(D_{\bar\jmath} \overline{\mathcal{W}})\epsilon_2 &=0\,,\\
3(\slashed \partial \alpha) \epsilon_1 + \f{i}4 \e^{K/2}(\partial_\alpha \overline{\mathcal{W}})\epsilon_2 &=0\,,\\
(\slashed \partial \beta) \epsilon_1 + \f{i}4 \e^{K/2}(\partial_\beta \overline{\mathcal{W}})\epsilon_2 &=0\,,\\
\end{split}
\ee
where
\be
Q_\mu = -i(\partial_\mu z^i\partial_i-\partial_\mu{\bar z}^{\bar\imath}\partial_{\bar\imath})K\,.
\ee

We now consider a series of two truncations of the ten-scalar model. The first one involves setting $\beta=0$ which is a $\U(1)$ invariant sector of the ten-scalar model. Since $\beta=0$ we must also have $\beta'=0$ which implies
\be
\partial_\beta \mathcal{W}=0 \quad\Leftrightarrow \quad (z_1-z_3)(z_4-z_2)=0\,,
\ee
and we can pick $z_4 = z_2$ without loss of generality. The other choice $z_3=z_1$ is related to the first by a discrete symmetry of the model. Hence setting $\beta=0$ truncates not just one scalar away, but three and we are left with
\be\label{7scalarmodel}
\begin{split}
K &= - \log(2\Im z_1)- 2\log(2\Im z_2)- \log(2\Im z_3)\,,\\
\mathcal{W} &=  g\Big(\e^{-4\alpha}(z_1z_3+z_2^2)+2z_2\e^{2\alpha}(z_1+z_3)\Big)\,.
\end{split}
\ee

Next we assume that $z_2$ is pure imaginary, and thus truncate to the $\U(1)\times\U(1)$ invariant six-scalar sector which is the main subject of this paper. To recover the six-scalar model as written in Section~\ref{6scalarmodel} we write
\be
z_2 = i \e^t\,,
\ee
with $t$ real. Inserting this into the seven-scalar K{\"a}hler potential and superpotential in \eqref{7scalarmodel} we recover the theory in \eqref{Kaehlerandsuperpot}.

\bibliographystyle{utphys}
\bibliography{references}

\end{document}